\documentclass[12pt]{article}
\usepackage{amssymb}
\begin{document}

\author{C. Bizdadea\thanks{%
e-mail address: bizdadea@central.ucv.ro}, C. C. Ciob\^\i rc\u {a}\thanks{%
e-mail address: ciobarca@central.ucv.ro}, E. M. Cioroianu\thanks{%
e-mail address: manache@central.ucv.ro}, \and S. O. Saliu\thanks{%
e-mail address: osaliu@central.ucv.ro}, S. C. S\u {a}raru\thanks{%
e-mail address: scsararu@central.ucv.ro} \\
Faculty of Physics, University of Craiova\\
13 A. I. Cuza Str., Craiova Ro-1100, Romania}
\title{Interactions of a massless tensor field with the mixed symmetry of the
Riemann tensor. No-go results}
\maketitle

\begin{abstract}
Non-trivial, consistent interactions of a free, massless tensor field $%
t_{\mu \nu |\alpha \beta }$ with the mixed symmetry of the Riemann tensor
are studied in the following cases: self-couplings, cross-interactions with
a Pauli-Fierz field and cross-couplings with purely matter theories. The
main results, obtained from BRST cohomological techniques under the
assumptions on smoothness, locality, Lorentz covariance and Poincar\'{e}
invariance of the deformations, combined with the requirement that the
interacting Lagrangian is at most second-order derivative, can be
synthesized into: no consistent self-couplings exist, but a
cosmological-like term; no cross-interactions with the Pauli-Fierz field can
be added; no non-trivial consistent cross-couplings with the matter theories
such that the matter fields gain gauge transformations are allowed.

\end{abstract}

\section{Introduction}

Mixed symmetry type tensor fields~\cite{curt,aul,labast,burd,zinov1} are
involved in many physically interesting theories, like superstrings,
supergravities or supersymmetric high spin theories. The study of gauge
theories with mixed symmetry type tensor fields revealed several issues,
like the dual formulation of field theories of spin two or higher~\cite
{dualsp1,dualsp2,dualsp2a,dualsp3,dualsp4,dualsp5}, the impossibility of
consistent cross-interactions in the dual formulation of linearized gravity~%
\cite{lingr} or a Lagrangian first-order approach~\cite{zinov2,zinov3} to
some classes of free massless mixed symmetry type tensor gauge fields,
suggestively resembling to the tetrad formalism of General Relativity. One
of the most important aspects related to this type of gauge models is the
analysis of their consistent interactions, among themselves, as well as with
higher-spin gauge theories~\cite{high1,high2,high3,high4,noijhep}. The best
approach to this matter is the cohomological one, based on the deformation
of the solution to the master equation~\cite{def}. The aim of our paper is
to investigate the manifestly covariant consistent interactions involving a
single, free, massless tensor gauge field $t_{\mu \nu |\alpha \beta }$, with
the mixed symmetry of the Riemann tensor, in three distinct situations:
self-couplings, interactions with the massless spin-two field (described in
the free limit by the Pauli-Fierz action~\cite{pf}), and couplings with
purely matter theories.

Our procedure relies on the deformation of the solution to the master
equation by means of local BRST cohomology. For each situation, we initially
determine the associated free antifield-BRST symmetry $s$, which splits as
the sum between the Koszul-Tate differential and the exterior longitudinal
derivative only, $s=\delta +\gamma $. Then, we solve the basic equations of
the deformation procedure. Under the supplementary assumptions on
smoothness, locality, Lorentz covariance and Poincar\'{e} invariance of the
deformations, as well as on the maximum derivative order of the interacting
Lagrangian being equal to two, we prove the following no-go results: (i) the
self-interactions of the tensor field with the mixed symmetry of the Riemann
tensor do not modify either the original gauge algebra or the gauge
transformations and, in fact, reduce to a cosmological-like term; (ii) there
are no consistent cross-interactions between such a tensor field and the
Pauli-Fierz model. Only the Pauli-Fierz theory leads to consistent
self-interactions, described by the Einstein-Hilbert action with a
cosmological term, invariant under diffeomorphisms; (iii) there are no
couplings with purely matter theories such that the matter fields become
endowed with gauge transformations.

The paper is organized in eight sections. Section \ref{2} is dedicated to
the Lagrangian formulation of the free massless tensor gauge field with the
mixed symmetry of the Riemann tensor, emphasizing its relationship with the
generalized 3-differential complex. In Section \ref{3} we construct the
associated BRST symmetry and in Section \ref{4} we briefly review the
antifield-BRST deformation procedure. The following three sections represent
the core of the paper and respectively address the problem of
self-interactions, interactions with the Pauli-Fierz field, and couplings
with purely matter fields. Section 8 ends the paper with the main
conclusions.

\section{Free model\label{2}}

\subsection{Field equations and gauge transformations\label{2.1}}

The starting point is given by the free Lagrangian action 
\begin{eqnarray}
S_{0}\left[ t_{\mu \nu |\alpha \beta }\right] &=&\int d^{D}x\left( \frac{1}{8%
}\left( \partial ^{\lambda }t^{\mu \nu |\alpha \beta }\right) \left(
\partial _{\lambda }t_{\mu \nu |\alpha \beta }\right) -\frac{1}{2}\left(
\partial _{\mu }t^{\mu \nu |\alpha \beta }\right) \left( \partial ^{\lambda
}t_{\lambda \nu |\alpha \beta }\right) \right.  \nonumber \\
&&-\left( \partial _{\mu }t^{\mu \nu |\alpha \beta }\right) \left( \partial
_{\beta }t_{\nu \alpha }\right) -\frac{1}{2}\left( \partial ^{\lambda
}t^{\nu \beta }\right) \left( \partial _{\lambda }t_{\nu \beta }\right) 
\nonumber \\
&&\left. +\left( \partial _{\nu }t^{\nu \beta }\right) \left( \partial
^{\lambda }t_{\lambda \beta }\right) -\frac{1}{2}\left( \partial _{\nu
}t^{\nu \beta }\right) \left( \partial _{\beta }t\right) +\frac{1}{8}\left(
\partial ^{\lambda }t\right) \left( \partial _{\lambda }t\right) \right) ,
\label{r1}
\end{eqnarray}
in a Minkowski-flat spacetime of dimension $D\geq 5$, endowed with a metric
tensor of `mostly plus' signature $\sigma _{\mu \nu }=\sigma ^{\mu \nu
}=\left( -++++\cdots \right) $. The massless tensor field $t_{\mu \nu
|\alpha \beta }$ of degree four has the mixed symmetry of the linearized
Riemann tensor, and hence transforms according to an irreducible
representation of $GL\left( D,\mathbb{R}\right) $, corresponding to a
rectangular Young diagram $\left( 2,2\right) $ with two columns and two
rows, so it is separately antisymmetric in the pairs $\left\{ \mu ,\nu
\right\} $ and $\left\{ \alpha ,\beta \right\} $, is symmetric under the
interchange of these pairs ($\left\{ \mu ,\nu \right\} \longleftrightarrow
\left\{ \alpha ,\beta \right\} $), $t_{\mu \nu |\alpha \beta }=t_{\alpha
\beta |\mu \nu }$, and satisfies the identity 
\begin{equation}
t_{\left[ \mu \nu |\alpha \right] \beta }\equiv 0  \label{r5}
\end{equation}
associated with the above diagram, which we will refer to as the Bianchi I
identity. Here and in the sequel the symbol $\left[ \mu \nu \cdots \right] $
denotes the operation of antisymmetrization with respect to the indices
between brackets, without normalization factors. (For instance, the
left-hand side of (\ref{r5}) contains only three terms $t_{\left[ \mu \nu
|\alpha \right] \beta }=t_{\mu \nu |\alpha \beta }+t_{\nu \alpha |\mu \beta
}+t_{\alpha \mu |\nu \beta }$.) The notation $t_{\nu \beta }$ signifies the
simple trace of the original tensor field, which is symmetric $t_{\nu \beta
}=\sigma ^{\mu \alpha }t_{\mu \nu |\alpha \beta }$, while $t$ denotes its
double trace, which is a scalar, $t=\sigma ^{\nu \beta }t_{\nu \beta }$. A
generating set of gauge transformations for the action (\ref{r1}) reads as 
\begin{equation}
\delta _{\epsilon }t_{\mu \nu |\alpha \beta }=\partial _{\mu }\epsilon
_{\alpha \beta |\nu }-\partial _{\nu }\epsilon _{\alpha \beta |\mu
}+\partial _{\alpha }\epsilon _{\mu \nu |\beta }-\partial _{\beta }\epsilon
_{\mu \nu |\alpha },  \label{r8}
\end{equation}
with the bosonic gauge parameters $\epsilon _{\mu \nu |\alpha }$
transforming according to an irreducible representation of $GL\left( D,%
\mathbb{R}\right) $, corresponding to a Young diagram $\left( 2,1\right) $
with two columns and two rows, being therefore antisymmetric in the pair $%
\left\{ \mu ,\nu \right\} $ and satisfying the identity 
\begin{equation}
\epsilon _{\left[ \mu \nu |\alpha \right] }\equiv 0.  \label{r10}
\end{equation}
The identity (\ref{r10}) is required in order to ensure that the gauge
transformations (\ref{r8}) check the same Bianchi I identity like the fields
themselves, namely, $\delta _{\epsilon }t_{\left[ \mu \nu |\alpha \right]
\beta }\equiv 0$. The above generating set of gauge transformations is
abelian and off-shell first-stage reducible since if we make the
transformation 
\begin{equation}
\epsilon _{\mu \nu |\alpha }=2\partial _{\alpha }\theta _{\mu \nu }-\partial
_{\left[ \mu \right. }\theta _{\left. \nu \right] \alpha },  \label{r12}
\end{equation}
with $\theta _{\mu \nu }$ an arbitrary antisymmetric tensor ($\theta _{\mu
\nu }=-\theta _{\nu \mu }$), then the gauge transformations of the tensor
field identically vanish, $\delta _{\epsilon }t_{\mu \nu |\alpha \beta
}\equiv 0$. In the meantime, the transformation (\ref{r12}) agrees with the
identity (\ref{r10}) checked by the gauge parameters.

The field equations resulting from the action (\ref{r1}) take the form 
\begin{equation}
\frac{\delta S_{0}}{\delta t^{\mu \nu |\alpha \beta }}\equiv -\frac{1}{4}%
T_{\mu \nu |\alpha \beta }\approx 0,  \label{r15}
\end{equation}
where 
\begin{eqnarray}
T_{\mu \nu |\alpha \beta } &=&\Box t_{\mu \nu |\alpha \beta }+\partial
^{\rho }\left( \partial _{\mu }t_{\alpha \beta |\nu \rho }-\partial _{\nu
}t_{\alpha \beta |\mu \rho }+\partial _{\alpha }t_{\mu \nu |\beta \rho
}-\partial _{\beta }t_{\mu \nu |\alpha \rho }\right)  \nonumber \\
&&+\left( \partial _{\mu }\partial _{\alpha }t_{\beta \nu }-\partial _{\mu
}\partial _{\beta }t_{\alpha \nu }-\partial _{\nu }\partial _{\alpha
}t_{\beta \mu }+\partial _{\nu }\partial _{\beta }t_{\alpha \mu }\right) 
\nonumber \\
&&-\frac{1}{2}\partial ^{\lambda }\partial ^{\rho }\left( \sigma _{\mu
\alpha }\left( t_{\lambda \beta |\nu \rho }+t_{\lambda \nu |\beta \rho
}\right) -\sigma _{\mu \beta }\left( t_{\lambda \alpha |\nu \rho
}+t_{\lambda \nu |\alpha \rho }\right) \right.  \nonumber \\
&&\left. -\sigma _{\nu \alpha }\left( t_{\lambda \beta |\mu \rho
}+t_{\lambda \mu |\beta \rho }\right) +\sigma _{\nu \beta }\left( t_{\lambda
\alpha |\mu \rho }+t_{\lambda \mu |\alpha \rho }\right) \right)  \nonumber \\
&&-\Box \left( \sigma _{\mu \alpha }t_{\beta \nu }-\sigma _{\mu \beta
}t_{\alpha \nu }-\sigma _{\nu \alpha }t_{\beta \mu }+\sigma _{\nu \beta
}t_{\alpha \mu }\right)  \nonumber \\
&&+\partial ^{\rho }\left( \sigma _{\mu \alpha }\left( \partial _{\beta
}t_{\nu \rho }+\partial _{\nu }t_{\beta \rho }\right) -\sigma _{\mu \beta
}\left( \partial _{\alpha }t_{\nu \rho }+\partial _{\nu }t_{\alpha \rho
}\right) \right.  \nonumber \\
&&\left. -\sigma _{\nu \alpha }\left( \partial _{\beta }t_{\mu \rho
}+\partial _{\mu }t_{\beta \rho }\right) +\sigma _{\nu \beta }\left(
\partial _{\alpha }t_{\mu \rho }+\partial _{\mu }t_{\alpha \rho }\right)
\right)  \nonumber \\
&&-\frac{1}{2}\left( \sigma _{\mu \alpha }\partial _{\beta }\partial _{\nu
}-\sigma _{\mu \beta }\partial _{\alpha }\partial _{\nu }-\sigma _{\nu
\alpha }\partial _{\beta }\partial _{\mu }+\sigma _{\nu \beta }\partial
_{\alpha }\partial _{\mu }\right) t  \nonumber \\
&&-\left( \sigma _{\mu \alpha }\sigma _{\nu \beta }-\sigma _{\mu \beta
}\sigma _{\nu \alpha }\right) \left( \partial ^{\lambda }\partial ^{\rho
}t_{\lambda \rho }-\frac{1}{2}\Box t\right) .  \label{r16}
\end{eqnarray}
Obviously, the tensor $T_{\mu \nu |\alpha \beta }$ displays the same mixed
symmetry properties like the tensor field $t_{\mu \nu |\alpha \beta }$. It
is useful to compute its simple and double traces 
\begin{eqnarray}
\sigma ^{\mu \alpha }T_{\mu \nu |\alpha \beta } &\equiv &T_{\nu \beta
}=\left( 4-D\right) \left( \frac{1}{2}\partial ^{\lambda }\partial ^{\rho
}\left( t_{\lambda \nu |\beta \rho }+t_{\lambda \beta |\nu \rho }\right)
\right.  \nonumber \\
&&+\Box t_{\nu \beta }-\partial ^{\rho }\left( \partial _{\nu }t_{\beta \rho
}+\partial _{\beta }t_{\nu \rho }\right)  \nonumber \\
&&\left. +\frac{1}{2}\partial _{\nu }\partial _{\beta }t+\sigma _{\nu \beta
}\left( \partial ^{\lambda }\partial ^{\rho }t_{\lambda \rho }-\frac{1}{2}%
\Box t\right) \right) ,  \label{r16a}
\end{eqnarray}
\begin{equation}
\sigma ^{\nu \beta }T_{\nu \beta }\equiv T=-\left( 4-D\right) \left(
3-D\right) \left( \partial ^{\lambda }\partial ^{\rho }t_{\lambda \rho }-%
\frac{1}{2}\Box t\right) .  \label{r16b}
\end{equation}
Obviously, its simple trace is a symmetric tensor, while its double trace is
a scalar. The gauge invariance of the Lagrangian action (\ref{r1}) under the
transformations (\ref{r8}) is equivalent to the fact that the functions
defining the field equations are not all independent, but rather obey the
Noether identities 
\begin{equation}
\partial ^{\mu }\frac{\delta S_{0}}{\delta t^{\mu \nu |\alpha \beta }}\equiv
-\frac{1}{4}\partial ^{\mu }T_{\mu \nu |\alpha \beta }=0,  \label{r17}
\end{equation}
while the first-stage reducibility shows that not all of the above Noether
identities are independent. It can be checked that the functions (\ref{r16})
defining the field equations, the gauge generators, as well as the
first-order reducibility functions, satisfy the general regularity
assumptions from~\cite{genreg}, such that the model under discussion is
described by a normal gauge theory of Cauchy order equal to three.

\subsection{Interpretation via the generalized 3-complex\label{2.2}}

This model describes a free gauge theory that can be interpreted in a
consistent manner in terms of the generalized differential complex~\cite
{genpoinc} $\Omega _{2}\left( \mathcal{M}\right) $ of tensor fields with
mixed symmetries corresponding to a maximal sequence of Young diagrams with
two columns, defined on a pseudo-Riemannian manifold $\mathcal{M}$ of
dimension $D$. Let us denote by $\bar{d}$ the associated operator
(3-differential) that is third-order nilpotent, $\bar{d}^{3}=0$, and by $%
\Omega _{2}^{p}\left( \mathcal{M}\right) $ the vector space spanned by the
tensor fields from $\Omega _{2}\left( \mathcal{M}\right) $ with $p$ entries.
The action of $\bar{d}$ on an element pertaining to $\Omega _{2}^{p}\left( 
\mathcal{M}\right) $ results in a tensor from $\Omega _{2}^{p+1}\left( 
\mathcal{M}\right) $ with one spacetime derivative, the action of $\bar{d}%
^{2}$ on a similar element leads to a tensor from $\Omega _{2}^{p+2}\left( 
\mathcal{M}\right) $ containing two spacetime derivatives, while the action
of $\bar{d}^{3}$ on any such element identically vanishes. In brief, the
generalized 3-complex $\Omega _{2}\left( \mathcal{M}\right) $ may
suggestively be represented through the commutative diagram 
\[
\begin{array}{ccccccccccc}
&  &  &  &  &  &  &  &  &  & \cdots \\ 
&  &  &  &  &  &  &  &  & \stackrel{\bar{d}^{2}}{\nearrow } &  \\ 
&  &  &  &  &  &  &  & \Omega _{2}^{8} & \stackrel{\bar{d}}{\rightarrow } & 
\cdots \\ 
&  &  &  &  &  &  & \stackrel{\bar{d}^{2}}{\nearrow } & \uparrow ^{\bar{d}}
&  &  \\ 
&  &  &  &  &  & \mathbf{\Omega }_{2}^{6} & \stackrel{\bar{d}}{\rightarrow }
& \mathbf{\Omega }_{2}^{7} &  &  \\ 
&  &  &  &  & \stackrel{\bar{d}^{2}}{\nearrow } & \uparrow ^{\bar{d}} &  & 
&  &  \\ 
&  &  &  & \mathbf{\Omega }_{2}^{4} & \stackrel{\bar{d}}{\rightarrow } & 
\mathbf{\Omega }_{2}^{5} &  &  &  &  \\ 
&  &  & \stackrel{\bar{d}^{2}}{\nearrow } & \uparrow ^{\bar{d}} &  &  &  & 
&  &  \\ 
&  & \Omega _{2}^{2} & \stackrel{\bar{d}}{\rightarrow } & \mathbf{\Omega }%
_{2}^{3} &  &  &  &  &  &  \\ 
& \stackrel{\bar{d}^{2}}{\nearrow } & \uparrow ^{\bar{d}} &  &  &  &  &  & 
&  &  \\ 
\Omega _{2}^{0} & \stackrel{\bar{d}}{\rightarrow } & \Omega _{2}^{1} &  &  & 
&  &  &  &  & 
\end{array}
\]
where the third-order nilpotency of $\bar{d}$ means that any vertical arrow
followed by the closest higher diagonal arrow maps to zero, and the same
with respect to any diagonal arrow followed by the closest higher horizontal
one. Its bold part emphasizes the sequences that apply to the model under
discussion: the first one governs the dynamics and indicates the presence of
some gauge symmetry 
\begin{equation}
\begin{array}{ccccc}
\Omega _{2}^{4} &  & \Omega _{2}^{6} &  & \Omega _{2}^{7} \\ 
\begin{array}{l}
\mathrm{field} \\ 
t_{\mu \nu |\alpha \beta }
\end{array}
& \stackrel{\bar{d}^{2}}{\rightarrow } & 
\begin{array}{l}
\mathrm{curvature} \\ 
F_{\mu \nu \lambda |\alpha \beta \gamma }
\end{array}
& \stackrel{\bar{d}}{\rightarrow } & 
\begin{array}{l}
\mathrm{Bianchi\;II} \\ 
\partial _{\left[ \rho \right. }F_{\left. \mu \nu \lambda \right] |\alpha
\beta \gamma }=0
\end{array}
\end{array}
,  \label{r18}
\end{equation}
while the second sequence solves the gauge symmetry 
\begin{equation}
\begin{array}{ccccc}
\Omega _{2}^{3} &  & \Omega _{2}^{4} &  & \Omega _{2}^{6} \\ 
\begin{array}{l}
\mathrm{gauge\;param.} \\ 
\epsilon _{\alpha \beta |\mu }
\end{array}
& \stackrel{\bar{d}}{\rightarrow } & 
\begin{array}{l}
\mathrm{gauge\;transf.} \\ 
\delta _{\epsilon }t_{\mu \nu |\alpha \beta }
\end{array}
& \stackrel{\bar{d}^{2}}{\rightarrow } & 
\begin{array}{l}
\mathrm{gauge\;inv.\;objects} \\ 
\delta _{\epsilon }F_{\mu \nu \lambda |\alpha \beta \gamma }=0
\end{array}
\end{array}
.  \label{r19}
\end{equation}
Let us discuss the previous sequences. Starting from the tensor field $%
t_{\mu \nu |\alpha \beta }$ from $\Omega _{2}^{4}$, we can construct its
curvature tensor $F_{\mu \nu \lambda |\alpha \beta \gamma }$, defined via 
\begin{eqnarray}
\left( \bar{d}^{2}t\right) _{\mu \nu \lambda \alpha \beta \gamma } &\sim
&F_{\mu \nu \lambda |\alpha \beta \gamma }=\partial _{\lambda }\partial
_{\gamma }t_{\mu \nu |\alpha \beta }+\partial _{\mu }\partial _{\gamma
}t_{\nu \lambda |\alpha \beta }+\partial _{\nu }\partial _{\gamma
}t_{\lambda \mu |\alpha \beta }  \nonumber \\
&&+\partial _{\lambda }\partial _{\alpha }t_{\mu \nu |\beta \gamma
}+\partial _{\mu }\partial _{\alpha }t_{\nu \lambda |\beta \gamma }+\partial
_{\nu }\partial _{\alpha }t_{\lambda \mu |\beta \gamma }  \nonumber \\
&&+\partial _{\lambda }\partial _{\beta }t_{\mu \nu |\gamma \alpha
}+\partial _{\mu }\partial _{\beta }t_{\nu \lambda |\gamma \alpha }+\partial
_{\nu }\partial _{\beta }t_{\lambda \mu |\gamma \alpha },  \label{curv}
\end{eqnarray}
which is second-order in the spacetime derivatives and belongs to $\Omega
_{2}^{6}$. Thus, the curvature tensor transforms in an irreducible
representation of $GL\left( D,\mathbb{R}\right) $ and exhibits the
symmetries of a rectangular two-column Young diagram $\left( 3,3\right) $,
being separately antisymmetric in the indices $\left\{ \mu ,\nu ,\lambda
\right\} $ and $\left\{ \alpha ,\beta ,\gamma \right\} $, symmetric under
the interchange $\left\{ \mu ,\nu ,\lambda \right\} \longleftrightarrow
\left\{ \alpha ,\beta ,\gamma \right\} $, and obeying the (algebraic)
Bianchi I identity 
\begin{equation}
F_{\left[ \mu \nu \lambda |\alpha \right] \beta \gamma }\equiv 0.
\label{r22}
\end{equation}
The action of $\bar{d}$ on $F_{\mu \nu \lambda |\alpha \beta \gamma }$ maps
to zero 
\begin{equation}
\left( \bar{d}^{3}t\right) _{\rho \mu \nu \lambda \alpha \beta \gamma
}=\left( \bar{d}F\right) _{\rho \mu \nu \lambda \alpha \beta \gamma }\sim
\partial _{\left[ \rho \right. }F_{\left. \mu \nu \lambda \right] |\alpha
\beta \gamma }\equiv 0,  \label{r23}
\end{equation}
and represents nothing but the (differential) Bianchi II identity for the
curvature tensor. Since the curvature and its traces are the most general
non-vanishing second-order derivative quantities in $\Omega _{2}\left( 
\mathcal{M}\right) $ constructed from $t_{\mu \nu |\alpha \beta }$, we
expect that the free field equations (\ref{r15}) completely rely on it. The
formula (\ref{r23}) shows that the corresponding field equations cannot be
all independent, but satisfy some Noether identities related to the Bianchi
II identity of the curvature tensor. This already points out that the
searched for free Lagrangian action must be invariant under a certain gauge
symmetry. The second sequence, namely (\ref{r19}), gives the form of the
gauge invariance. As the free field equations involve $F_{\mu \nu \lambda
|\alpha \beta \gamma }$, it is natural to require that these are the most
general gauge invariant quantities 
\begin{equation}
\delta _{\epsilon }\left( \bar{d}^{2}t\right) _{\mu \nu \lambda \alpha \beta
\gamma }\sim \delta _{\epsilon }F_{\mu \nu \lambda |\alpha \beta \gamma }=0.
\label{r24}
\end{equation}
This matter is immediately solved if we take 
\begin{equation}
\left( \bar{d}\epsilon \right) _{\mu \nu \alpha \beta }\sim \partial _{\mu
}\epsilon _{\alpha \beta |\nu }-\partial _{\nu }\epsilon _{\alpha \beta |\mu
}+\partial _{\alpha }\epsilon _{\mu \nu |\beta }-\partial _{\beta }\epsilon
_{\mu \nu |\alpha }=\delta _{\epsilon }t_{\mu \nu |\alpha \beta },
\label{r25}
\end{equation}
where the gauge parameters $\epsilon _{\mu \nu |\alpha }$ pertain to $\Omega
_{2}^{3}$, because, on account of the third-order nilpotency of $\bar{d}$,
we find that 
\begin{equation}
\delta _{\epsilon }F_{\mu \nu \lambda |\alpha \beta \gamma }\sim \left( \bar{%
d}^{3}\epsilon \right) _{\mu \nu \lambda \alpha \beta \gamma }\equiv 0.
\label{r26}
\end{equation}
Clearly, the relation (\ref{r25}) coincides with the gauge transformations (%
\ref{r8}).

We complete our discussion by exemplifying the construction of the free
field equations. Let us denote by $S_{0}^{\prime }\left[ t_{\mu \nu |\alpha
\beta }\right] $ a free, second-order derivative action that is gauge
invariant under (\ref{r25}), and by $\delta S_{0}^{\prime }/\delta t^{\mu
\nu |\alpha \beta }$ its functional derivatives with respect to the fields,
which are imposed to depend linearly on the undifferentiated curvature
tensor. Then, as these functional derivatives must have the same mixed
symmetry like $t_{\mu \nu |\alpha \beta }$, it follows that they necessarily
determine a tensor from $\Omega _{2}^{4}$. The operations that can be
performed with respect to the curvature tensor in order to reduce its number
of indices without increasing its derivative order is to take its simple,
double, and respectively, triple traces 
\begin{eqnarray}
F_{\mu \nu |\alpha \beta } &=&\sigma ^{\lambda \gamma }F_{\mu \nu \lambda
|\alpha \beta \gamma }\in \Omega _{2}^{4},  \label{r28} \\
F_{\mu \alpha } &=&\sigma ^{\nu \beta }F_{\mu \nu |\alpha \beta }\in \Omega
_{2}^{2},  \label{r29} \\
F &=&\sigma ^{\mu \alpha }F_{\mu \alpha }\in \Omega _{2}^{0},  \label{r30}
\end{eqnarray}
where $F_{\mu \alpha }$ is symmetric and $F$ is a scalar. The only
combinations formed with these quantities that belong to $\Omega _{2}^{4}$
are generated by 
\begin{equation}
F_{\mu \nu |\alpha \beta },  \label{r30a}
\end{equation}
\begin{equation}
\sigma _{\mu \alpha }F_{\beta \nu }-\sigma _{\mu \beta }F_{\alpha \nu
}-\sigma _{\nu \alpha }F_{\beta \mu }+\sigma _{\nu \beta }F_{\alpha \mu },
\label{r30b}
\end{equation}
and 
\begin{equation}
\left( \sigma _{\mu \alpha }\sigma _{\nu \beta }-\sigma _{\mu \beta }\sigma
_{\nu \alpha }\right) F,  \label{r31}
\end{equation}
so in principle $\delta S_{0}^{\prime }/\delta t^{\mu \nu |\alpha \beta }$
can be written as a linear combination of (\ref{r30a}--\ref{r31}) with
coefficients that are real constants. However, the requirement that the
above linear combination indeed stands for the functional derivatives of a
sole functional restricts the parametrization of the functional derivatives,
and therefore of the Lagrangian action, by means of one constant only 
\begin{eqnarray}
\delta S_{0}^{\prime }/\delta t^{\mu \nu |\alpha \beta } &=&\lambda \left(
F_{\mu \nu |\alpha \beta }-\frac{1}{2}\left( \sigma _{\mu \alpha }F_{\beta
\nu }-\sigma _{\mu \beta }F_{\alpha \nu }-\sigma _{\nu \alpha }F_{\beta \mu
}+\sigma _{\nu \beta }F_{\alpha \mu }\right) \right.  \nonumber \\
&&\left. +\frac{1}{6}\left( \sigma _{\mu \alpha }\sigma _{\nu \beta }-\sigma
_{\mu \beta }\sigma _{\nu \alpha }\right) F\right) .  \label{r33}
\end{eqnarray}
If in (\ref{r33}) we take the particular value 
\begin{equation}
\lambda =-\frac{1}{4},  \label{r34}
\end{equation}
we recover the Lagrangian action (\ref{r1}) together with the field
equations (\ref{r15}). This also allows us to identify the expression of $%
T_{\mu \nu |\alpha \beta }$ from (\ref{r15}--\ref{r16}) in terms of the
curvature tensor like 
\begin{eqnarray}
T_{\mu \nu |\alpha \beta } &=&\left( F_{\mu \nu |\alpha \beta }-\frac{1}{2}%
\left( \sigma _{\mu \alpha }F_{\beta \nu }-\sigma _{\mu \beta }F_{\alpha \nu
}-\sigma _{\nu \alpha }F_{\beta \mu }+\sigma _{\nu \beta }F_{\alpha \mu
}\right) \right.  \nonumber \\
&&\left. +\frac{1}{6}\left( \sigma _{\mu \alpha }\sigma _{\nu \beta }-\sigma
_{\mu \beta }\sigma _{\nu \alpha }\right) F\right) .  \label{r35}
\end{eqnarray}

At this point, we can easily see the relationship of the field equations (%
\ref{r15}) and their Noether identities (\ref{r17}) with the curvature
tensor (\ref{curv}) and accompanying Bianchi II identity (\ref{r23}). First,
we observe that the field equations (\ref{r15}) are completely equivalent
with the vanishing of the simple trace of the curvature tensor 
\begin{equation}
T_{\mu \nu |\alpha \beta }\approx 0\Longleftrightarrow F_{\mu \nu |\alpha
\beta }\approx 0.  \label{r36}
\end{equation}
The direct statement holds due to the fact that $T_{\mu \nu |\alpha \beta }$
is expressed only through $F_{\mu \nu |\alpha \beta }$ and its traces, such
that its vanishing implies $F_{\mu \nu |\alpha \beta }\approx 0$. The
converse implication holds because the vanishing of the second and
respectively third component in the right-hand side of (\ref{r35}) is a
simple consequence of $F_{\mu \nu |\alpha \beta }\approx 0$. Second, the
Noether identities (\ref{r17}) are a direct consequence of the Bianchi II
identity for the curvature tensor 
\begin{equation}
\partial _{\left[ \mu \right. }F_{\left. \alpha \beta \lambda \right] |\nu
\rho \theta }\equiv 0\Rightarrow \partial ^{\mu }T_{\mu \nu |\alpha \beta
}\equiv 0.  \label{r37}
\end{equation}
Indeed, on the one hand the relation (\ref{r35}) yields 
\begin{equation}
\partial ^{\mu }T_{\mu \nu |\alpha \beta }=\partial ^{\mu }F_{\mu \nu
|\alpha \beta }-\frac{1}{2}\partial _{\left[ \alpha \right. }F_{\left. \beta
\right] \nu }+\frac{1}{2}\sigma _{\nu \left[ \alpha \right. }\left( \partial
^{\mu }F_{\left. \beta \right] \mu }-\frac{1}{3}\partial _{\left. \beta
\right] }F\right) .  \label{r38}
\end{equation}
On the other hand, simple computation leads to 
\begin{equation}
\sigma ^{\lambda \theta }\sigma ^{\mu \rho }\partial _{\left[ \mu \right.
}F_{\left. \alpha \beta \lambda \right] |\nu \rho \theta }=-2\left( \partial
^{\mu }F_{\mu \nu |\alpha \beta }-\frac{1}{2}\partial _{\left[ \alpha
\right. }F_{\left. \beta \right] \nu }\right) ,  \label{r39}
\end{equation}
\begin{equation}
2\sigma ^{\nu \beta }\left( \partial ^{\mu }F_{\mu \nu |\alpha \beta }-\frac{%
1}{2}\partial _{\left[ \alpha \right. }F_{\left. \beta \right] \nu }\right)
=3\left( \partial ^{\mu }F_{\alpha \mu }-\frac{1}{3}\partial _{\alpha
}F\right) .  \label{r40}
\end{equation}
Thus, according to (\ref{r39}--\ref{r40}), we can state that the Bianchi II
identity for the curvature tensor implies the identically vanishing of the
right-hand side of (\ref{r38}), and hence enforces the Noether identities (%
\ref{r17}) for the action (\ref{r1}).

Next, we point out the relation between the generalized cohomology of the
3-complex $\Omega _{2}\left( \mathcal{M}\right) $ and our model. The
generalized cohomology of the 3-complex $\Omega _{2}\left( \mathcal{M}%
\right) $ is given by the family of graded vector spaces $H_{k}\left( \bar{d}%
\right) =Ker\left( \bar{d}^{k}\right) /Im\left( \bar{d}^{3-k}\right) $, with 
$k=1,2$. Each vector space $H_{k}\left( \bar{d}\right) $ splits into the
cohomology spaces $H_{\left( k\right) }^{p}\left( \Omega _{2}\left( \mathcal{%
M}\right) \right) $, defined like the equivalence classes of tensors from $%
\Omega _{2}^{p}\left( \mathcal{M}\right) $ that are $\bar{d}^{k}$-closed,
with any two such tensors that differ by a $\bar{d}^{3-k}$-exact element in
the same equivalence class. The spaces $H_{\left( k\right) }^{p}$ are not
empty in general, even if $\mathcal{M}$ has a trivial topology. However, in
the case where $\mathcal{M}$ (assumed to be of dimension $D$) has the
topology of $\mathbb{R}^{D}$, the generalized Poincar\'{e} lemma~\cite
{genpoinc} applied to our situation states that the generalized cohomology
of the 3-differential $\bar{d}$ on tensors represented by rectangular
diagrams with two columns is empty in the space $\Omega _{2}\left( \mathbb{R}%
^{D}\right) $ of maximal two-column tensors, $H_{\left( k\right)
}^{2n}\left( \Omega _{2}\left( \mathbb{R}^{D}\right) \right) =0$, for $1\leq
n\leq D-1$ and $k=1,2$. In particular, for $n=3$ and $k=1$ we find that $%
H_{\left( 1\right) }^{6}\left( \Omega _{2}\left( \mathbb{R}^{D}\right)
\right) =0$ and thus, if the tensor $F_{\mu \nu \lambda |\alpha \beta \gamma
}$ with the mixed symmetry of the curvature tensor is $\bar{d}$-closed, then
it is also $\bar{d}^{2}$-exact. To put it otherwise, if this tensor
satisfies the Bianchi II identity $\partial _{\left[ \rho \right. }F_{\left.
\mu \nu \lambda \right] |\alpha \beta \gamma }\equiv 0$, then there exists
an element $t_{\mu \nu |\alpha \beta }$ with the mixed symmetry $\left( 2,2
\right) $, with the help of which $F_{\mu \nu \lambda |\alpha \beta \gamma } 
$ can precisely be written like in (\ref{curv}).

Finally, we observe that the formula (\ref{r35}) relates the functions
defining the free field equations (\ref{r15}) to the curvature tensor by a
generalized Hodge-duality. The generalized cohomology of $\bar{d}$ on $%
\Omega _{2}\left( \mathcal{M}\right) $ when $\mathcal{M}$ has the trivial
topology of $\mathbb{R}^{D}$ together with this type of generalized
Hodge-duality reveal many important features of the free model under study.
For example, if $\bar{T}_{\mu \nu |\alpha \beta }$ is a covariant tensor
field with the mixed symmetry $\left( 2,2 \right) $ and satisfies the
equations 
\begin{equation}
\partial ^{\mu }\bar{T}_{\mu \nu |\alpha \beta }=0,  \label{r40a}
\end{equation}
then there exists a tensor $\bar{\Phi}_{\mu \nu \rho |\alpha \beta \gamma
}\in \Omega _{2}\left( \mathbb{R}^{D}\right) $ with the mixed symmetry of
the curvature tensor, in terms of which 
\begin{equation}
\bar{T}_{\mu \nu |\alpha \beta }=\partial ^{\rho }\partial ^{\gamma }\bar{%
\Phi}_{\mu \nu \rho |\alpha \beta \gamma }+c\left( \sigma _{\mu \alpha
}\sigma _{\nu \beta }-\sigma _{\mu \beta }\sigma _{\nu \alpha }\right) ,
\label{r40b}
\end{equation}
with $c$ an arbitrary real constant. It is easy to check the above statement
in connection with the functions (\ref{r16}) that define the field equations
for the model under consideration. Indeed, direct computation provides $c=0$
and 
\begin{equation}
T_{\mu \nu |\alpha \beta }=\frac{1}{2}\partial ^{\rho }\partial ^{\gamma
}\Phi _{\mu \nu \rho |\alpha \beta \gamma },  \label{r40c}
\end{equation}
where 
\begin{eqnarray}
\Phi _{\mu \nu \rho |\alpha \beta \gamma } &=&\sigma _{\gamma \left[ \rho
\right. }t_{\left. \mu \nu \right] |\alpha \beta }+\sigma _{\alpha \left[
\rho \right. }t_{\left. \mu \nu \right] |\beta \gamma }+\sigma _{\beta
\left[ \rho \right. }t_{\left. \mu \nu \right] |\gamma \alpha }+\sigma
_{\rho \left[ \gamma \right. }t_{\left. \alpha \beta \right] |\mu \nu } 
\nonumber \\
&&+\sigma _{\mu \left[ \gamma \right. }t_{\left. \alpha \beta \right] |\nu
\rho }+\sigma _{\nu \left[ \gamma \right. }t_{\left. \alpha \beta \right]
|\rho \mu }-2\left( \sigma _{\gamma \left[ \rho \right. }\sigma _{\left. \mu
\right] \alpha }t_{\beta \nu }\right.  \nonumber \\
&&+\sigma _{\gamma \left[ \mu \right. }\sigma _{\left. \nu \right] \alpha
}t_{\beta \rho }+\sigma _{\gamma \left[ \nu \right. }\sigma _{\left. \rho
\right] \alpha }t_{\beta \mu }+\sigma _{\alpha \left[ \rho \right. }\sigma
_{\left. \mu \right] \beta }t_{\gamma \nu }  \nonumber \\
&&+\sigma _{\alpha \left[ \mu \right. }\sigma _{\left. \nu \right] \beta
}t_{\gamma \rho }+\sigma _{\alpha \left[ \nu \right. }\sigma _{\left. \rho
\right] \beta }t_{\gamma \mu }+\sigma _{\beta \left[ \rho \right. }\sigma
_{\left. \mu \right] \gamma }t_{\alpha \nu }  \nonumber \\
&&\left. +\sigma _{\beta \left[ \mu \right. }\sigma _{\left. \nu \right]
\gamma }t_{\alpha \rho }+\sigma _{\beta \left[ \nu \right. }\sigma _{\left.
\rho \right] \gamma }t_{\alpha \mu }\right) +\left( \sigma _{\gamma \left[
\rho \right. }\sigma _{\left. \mu \right] \alpha }\sigma _{\beta \nu }\right.
\nonumber \\
&&\left. +\sigma _{\gamma \left[ \mu \right. }\sigma _{\left. \nu \right]
\alpha }\sigma _{\beta \rho }+\sigma _{\gamma \left[ \nu \right. }\sigma
_{\left. \rho \right] \alpha }\sigma _{\beta \mu }\right) t,  \label{r40d}
\end{eqnarray}
such that the corresponding $\Phi _{\mu \nu \rho |\alpha \beta \gamma }$
indeed displays the mixed symmetry of the curvature tensor.

\section{Free BRST symmetry\label{3}}

In agreement with the general setting of the antibracket-antifield
formalism, the construction of the BRST symmetry for the free theory under
consideration starts with the identification of the BRST algebra on which
the BRST differential $s$ acts. The generators of the BRST algebra are of
two kinds: fields/ghosts and antifields. The ghost spectrum for the model
under study comprises the fermionic ghosts $\eta _{\alpha \beta |\mu }$
associated with the gauge parameters $\epsilon _{\alpha \beta |\mu }$ from (%
\ref{r8}), as well as the bosonic ghosts for ghosts $C_{\mu \nu }$ due to
the first-stage reducibility parameters $\theta _{\mu \nu }$ in (\ref{r12}).
In order to make compatible the behaviour of $\epsilon _{\alpha \beta |\mu }$
and $\theta _{\mu \nu }$ with that of the corresponding ghosts, we ask that $%
\eta _{\alpha \beta |\mu }$ satisfy the same properties like the gauge
parameters 
\begin{equation}
\eta _{\mu \nu |\alpha }=-\eta _{\nu \mu |\alpha },\;\eta _{\left[ \mu \nu
|\alpha \right] }\equiv 0  \label{r41}
\end{equation}
and that $C_{\mu \nu }$ is antisymmetric. The antifield spectrum is
organized into the antifields $t^{*\mu \nu |\alpha \beta }$ of the original
tensor field and those of the ghosts, $\eta ^{*\mu \nu |\alpha }$ and $%
C^{*\mu \nu }$, of statistics opposite to that of the associated
fields/ghosts. It is understood that $t^{*\mu \nu |\alpha \beta }$ is
subject to some conditions similar to those satisfied by the tensor field 
\begin{equation}
t^{*\mu \nu |\alpha \beta }=-t^{*\nu \mu |\alpha \beta }=-t^{*\mu \nu |\beta
\alpha }=t^{*\alpha \beta |\mu \nu },\;t^{*\left[ \mu \nu |\alpha \right]
\beta }\equiv 0  \label{r43}
\end{equation}
and, along the same line, it is required that 
\begin{equation}
\eta ^{*\mu \nu |\alpha }=-\eta ^{*\nu \mu |\alpha },\;\eta ^{*\left[ \mu
\nu |\alpha \right] }\equiv 0,\;C^{*\mu \nu }=-C^{*\nu \mu }.  \label{r44}
\end{equation}
We will denote the simple and double traces of $t^{*\mu \nu |\alpha \beta }$
by 
\begin{equation}
t^{*\nu \beta }=\sigma _{\mu \alpha }t^{*\mu \nu |\alpha \beta },\;t^{*\nu
\beta }=t^{*\beta \nu },\;t^{*}=\sigma _{\nu \beta }t^{*\nu \beta }.
\label{r44a}
\end{equation}

As both the gauge generators and reducibility functions for this model are
field-independent, it follows that the associated BRST differential ($%
s^{2}=0 $) splits into 
\begin{equation}
s=\delta +\gamma ,  \label{r45}
\end{equation}
where $\delta $ represents the Koszul-Tate differential ($\delta ^{2}=0$),
graded by the antighost number $\mathrm{agh}$ ($\mathrm{agh}\left( \delta
\right) =-1$), while $\gamma $ stands for the exterior derivative along the
gauge orbits and turns out to be a true differential ($\gamma ^{2}=0$) that
anticommutes with $\delta $ ($\delta \gamma +\gamma \delta =0$), whose
degree is named pure ghost number $\mathrm{pgh}$ ($\mathrm{pgh}\left( \gamma
\right) =1$). These two degrees do not interfere ($\mathrm{agh}\left( \gamma
\right) =0$, $\mathrm{pgh}\left( \delta \right) =0$). The overall degree
that grades the BRST differential is known as the ghost number ($\mathrm{gh}$%
) and is defined like the difference between the pure ghost number and the
antighost number, such that $\mathrm{gh}\left( s\right) =\mathrm{gh}\left(
\delta \right) =\mathrm{gh}\left( \gamma \right) =1$. According to the
standard rules of the BRST method, the corresponding degrees of the
generators from the BRST complex are valued like 
\begin{eqnarray}
\mathrm{pgh}\left( t_{\mu \nu |\alpha \beta }\right) &=&0,\;\mathrm{pgh}%
\left( \eta _{\mu \nu |\alpha }\right) =1,\;\mathrm{pgh}\left( C_{\mu \nu
}\right) =2,  \label{r46} \\
\mathrm{pgh}\left( t^{*\mu \nu |\alpha \beta }\right) &=&\mathrm{pgh}\left(
\eta ^{*\mu \nu |\alpha }\right) =\mathrm{pgh}\left( C^{*\mu \nu }\right) =0,
\label{r47} \\
\mathrm{agh}\left( t_{\mu \nu |\alpha \beta }\right) &=&\mathrm{agh}\left(
\eta _{\mu \nu |\alpha }\right) =\mathrm{agh}\left( C_{\mu \nu }\right) =0,
\label{r48} \\
\mathrm{agh}\left( t^{*\mu \nu |\alpha \beta }\right) &=&1,\;\mathrm{agh}%
\left( \eta ^{*\mu \nu |\alpha }\right) =2,\;\mathrm{agh}\left( C^{*\mu \nu
}\right) =3  \label{r49}
\end{eqnarray}
and the actions of $\delta $ and $\gamma $ on them are given by 
\begin{equation}
\gamma t_{\mu \nu |\alpha \beta }=\partial _{\mu }\eta _{\alpha \beta |\nu
}-\partial _{\nu }\eta _{\alpha \beta |\mu }+\partial _{\alpha }\eta _{\mu
\nu |\beta }-\partial _{\beta }\eta _{\mu \nu |\alpha },  \label{r50}
\end{equation}
\begin{equation}
\gamma \eta _{\mu \nu |\alpha }=2\partial _{\alpha }C_{\mu \nu }-\partial
_{\left[ \mu \right. }C_{\left. \nu \right] \alpha },\;\gamma C_{\mu \nu }=0,
\label{r51}
\end{equation}
\begin{equation}
\gamma t^{*\mu \nu |\alpha \beta }=0,\;\gamma \eta ^{*\mu \nu |\alpha
}=0,\;\gamma C^{*\mu \nu }=0,  \label{r54}
\end{equation}
\begin{equation}
\delta t_{\mu \nu |\alpha \beta }=0,\;\delta \eta _{\mu \nu |\alpha
}=0,\;\delta C_{\mu \nu }=0,  \label{r52}
\end{equation}
\begin{equation}
\delta t^{*\mu \nu |\alpha \beta }=\frac{1}{4}T^{\mu \nu |\alpha \beta
},\;\delta \eta ^{*\alpha \beta |\nu }=-4\partial _{\mu }t^{*\mu \nu |\alpha
\beta },\;\delta C^{*\mu \nu }=3\partial _{\alpha }\eta ^{*\mu \nu |\alpha },
\label{r53}
\end{equation}
with $T_{\mu \nu |\alpha \beta }$ expressed in (\ref{r16}) and both $\delta $
and $\gamma $ taken to act like right derivations.

The antifield-BRST differential is known to admit a canonical action in a
structure named antibracket and defined by decreeing the fields/ghosts
conjugated with the corresponding antifields, $s\cdot =\left( \cdot
,S\right) $, where $\left( ,\right) $ signifies the antibracket and $S$
denotes the canonical generator of the BRST symmetry. It is a bosonic
functional of ghost number zero involving both the field/ghost and antifield
spectra, which obeys the classical master equation 
\begin{equation}
\left( S,S\right) =0.  \label{r56}
\end{equation}
The classical master equation is equivalent with the second-order nilpotency
of $s$, $s^{2}=0$, while its solution encodes the entire gauge structure of
the associated theory. Taking into account the formulas (\ref{r50}--\ref{r53}%
), as well as the actions of $\delta $ and $\gamma $ in canonical form, we
find that the complete solution to the master equation for the model under
study reads as 
\begin{eqnarray}
S &=&S_{0}\left[ t_{\mu \nu |\alpha \beta }\right] +\int d^{D}x\left(
t^{*\mu \nu |\alpha \beta }\left( \partial _{\mu }\eta _{\alpha \beta |\nu
}-\partial _{\nu }\eta _{\alpha \beta |\mu }\right. \right.  \nonumber \\
&&\left. \left. +\partial _{\alpha }\eta _{\mu \nu |\beta }-\partial _{\beta
}\eta _{\mu \nu |\alpha }\right) +\eta ^{*\mu \nu |\alpha }\left( 2\partial
_{\alpha }C_{\mu \nu }-\partial _{\left[ \mu \right. }C_{\left. \nu \right]
\alpha }\right) \right) .  \label{r55}
\end{eqnarray}
The main ingredients of the antifield-BRST symmetry derived in this section
will be useful in the sequel at the analysis of consistent interactions that
can be added to the action (\ref{r1}) without changing its number of
independent gauge symmetries.

\section{Brief review of the antifield-BRST deformation procedure\label{4}}

There are three main types of consistent interactions that can be added to a
given gauge theory: \textit{(i)} the first type deforms only the Lagrangian
action, but not its gauge transformations, \textit{(ii)} the second kind
modifies both the action and its transformations, but not the gauge algebra,
and \textit{(iii)} the third, and certainly most interesting category,
changes everything, namely, the action, its gauge symmetries and the
accompanying algebra.

The reformulation of the problem of consistent deformations of a given
action and of its gauge symmetries in the antifield-BRST setting is based on
the observation that if a deformation of the classical theory can be
consistently constructed, then the solution to the master equation for the
initial theory can be deformed into 
\begin{equation}
\bar{S}=S+gS_{1}+g^{2}S_{2}+O\left( g^{3}\right) ,\;\varepsilon \left( \bar{S%
}\right) =0,\;\mathrm{gh}\left( \bar{S}\right) =0,  \label{r57}
\end{equation}
such that 
\begin{equation}
\left( \bar{S},\bar{S}\right) =0.  \label{r58}
\end{equation}
Here and in the sequel $\varepsilon \left( F\right) $ denotes the Grassmann
parity of $F$. The projection of (\ref{r58}) on the various powers in the
coupling constant induces the following tower of equations: 
\begin{eqnarray}
g^{0} &:&\left( S,S\right) =0,  \label{r59} \\
g^{1} &:&\left( S_{1},S\right) =0,  \label{r60} \\
g^{2} &:&\frac{1}{2}\left( S_{1},S_{1}\right) +\left( S_{2},S\right) =0,
\label{r61} \\
&&\vdots  \nonumber
\end{eqnarray}
The first equation is satisfied by hypothesis. The second one governs the
first-order deformation of the solution to the master equation ($S_{1}$) and
it shows that $S_{1}$ is a BRST co-cycle, $sS_{1}=0$, and hence it exists
and is local. The remaining equations are responsible for the higher-order
deformations of the solution to the master equation. No obstructions arise
in finding solutions to them as long as no further restrictions, such as
spacetime locality, are imposed. Obviously, only non-trivial first-order
deformations should be considered, since trivial ones ($S_{1}=sB$) lead to
trivial deformations of the initial theory and can be eliminated by
convenient redefinitions of the fields. Ignoring the trivial deformations,
it follows that $S_{1}$ is a non-trivial BRST-observable, $S_{1}\in
H^{0}\left( s\right) $. Once that the deformation equations (\ref{r60}--\ref
{r61}), etc., have been solved by means of specific cohomological
techniques, from the consistent non-trivial deformed solution to the master
equation we can extract all the information on the gauge structure of the
accompanying interacting theory.

\section{Self-interactions\label{5}}

The first task of our paper is to study the consistent interactions that can
be added to the free action (\ref{r1}) by means of solving the main
deformation equations, namely, (\ref{r60}--\ref{r61}), etc. For obvious
reasons, we consider only smooth, local, Lorentz-covariant and
Poincar\'{e}-invariant deformations. If we make the notation $S_{1}=\int
d^{D}x\,a$, with $a$ a local function, then the local form of the equation (%
\ref{r60}), which we have seen that controls the first-order deformation of
the solution to the master equation, becomes 
\begin{equation}
sa=\partial _{\mu }m^{\mu },\;\mathrm{gh}\left( a\right) =0,\;\varepsilon
\left( a\right) =0,  \label{r62}
\end{equation}
for some $m^{\mu }$, and it shows that the non-integrated density of the
first-order deformation pertains to the local cohomology of $s$ at ghost
number zero, $a\in H^{0}\left( s|d\right) $, where $d$ denotes the exterior
spacetime differential. In order to analyze the above equation, we develop $%
a $ according to the antighost number 
\begin{equation}
a=\sum\limits_{k=0}^{I}a_{k},\;\mathrm{agh}\left( a_{k}\right) =k,\;\mathrm{%
gh}\left( a_{k}\right) =0,\;\varepsilon \left( a_{k}\right) =0,  \label{r63}
\end{equation}
and assume, without loss of generality, that $a$ stops at some finite value $%
I$ of the antighost number.\footnote{%
This can be shown, for instance, like in~\cite{gen2} (Section 3), under the
sole assumption that the interacting Lagrangian at the first order in the
coupling constant, $a_{0}$, has a finite, but otherwise arbitrary derivative
order.} By taking into account the decomposition (\ref{r45}) of the BRST
differential, the equation (\ref{r62}) is equivalent to a tower of local
equations, corresponding to the various decreasing values of the antighost
number 
\begin{eqnarray}
\gamma a_{I} &=&\partial _{\mu }\stackrel{(I)}{m}^{\mu },  \label{r64} \\
\delta a_{I}+\gamma a_{I-1} &=&\partial _{\mu }\stackrel{(I-1)}{m}^{\mu },
\label{r65} \\
\delta a_{k}+\gamma a_{k-1} &=&\partial _{\mu }\stackrel{(k-1)}{m}^{\mu
},\;I-1\geq k\geq 1,  \label{r66}
\end{eqnarray}
where $\left( \stackrel{(k)}{m}^{\mu }\right) _{k=\overline{0,I}}$ are some
local currents, with $\mathrm{agh}\left( \stackrel{(k)}{m}^{\mu }\right) =k$%
. It can be proved\footnote{%
The proof is be given in Corollary 3.1 from \cite{0402099}.} that one can
replace the equation (\ref{r64}) at strictly positive antighost numbers with 
\begin{equation}
\gamma a_{I}=0,\;I>0.  \label{r67}
\end{equation}
In conclusion, under the assumption that $I>0$, the representative of
highest antighost number from the non-integrated density of the first-order
deformation can always be taken to be $\gamma $-closed, such that the
equation (\ref{r62}) associated with the local form of the first-order
deformation is completely equivalent to the tower of equations (\ref{r67})
and (\ref{r65}--\ref{r66}).

Before proceeding to the analysis of the solutions to the first-order
deformation equations, we briefly comment on the uniqueness and triviality
of such solutions. Due to the second-order nilpotency of $\gamma $ ($\gamma
^{2}=0$), the solution to the top equation (\ref{r67}) is clearly unique up
to $\gamma $-exact contributions, 
\begin{equation}
a_{I}\rightarrow a_{I}+\gamma b_{I},\;\mathrm{agh}\left( b_{I}\right) =I,\;%
\mathrm{pgh}\left( b_{I}\right) =I-1,\;\varepsilon \left( b_{I}\right) =1.
\label{r68}
\end{equation}
Meanwhile, if it turns out that $a_{I}$ reduces to $\gamma $-exact terms
only, $a_{I}=\gamma b_{I}$, then it can be made to vanish, $a_{I}=0$. In
other words, the non-triviality of the first-order deformation $a$ is
translated at its highest antighost number component into the requirement
that 
\begin{equation}
a_{I}\in H^{I}\left( \gamma \right) ,  \label{r69}
\end{equation}
where $H^{I}\left( \gamma \right) $ denotes the cohomology of the exterior
longitudinal derivative $\gamma $ at pure ghost number equal to $I$. At the
same time, the general condition on the non-integrated density of the
first-order deformation to be in a non-trivial cohomological class of $%
H^{0}\left( s|d\right) $ shows on the one hand that the solution to (\ref
{r62}) is unique up to $s$-exact pieces plus total divergences 
\begin{equation}
a\rightarrow a+sb+\partial _{\mu }n^{\mu },\;\mathrm{gh}\left( b\right)
=-1,\;\varepsilon \left( b\right) =1,\;\mathrm{gh}\left( n^{\mu }\right)
=0,\;\varepsilon \left( n^{\mu }\right) =0  \label{r73}
\end{equation}
and on the other hand that if the general solution to (\ref{r62}) is found
to be completely trivial, $a=sb+\partial _{\mu }n^{\mu }$, then it can be
made to vanish, $a=0$.

In the light of the above discussion, we pass to the investigation of the
solutions to the equations (\ref{r67}) and (\ref{r65}--\ref{r66}). We have
seen that $a_{I}$ belongs to the cohomology of the exterior longitudinal
derivative (see the formula \ref{r69})), such that we need to compute $%
H\left( \gamma \right) $ in order to construct the component of highest
antighost number from the first-order deformation. This matter is solved
with the help of the definitions (\ref{r50}--\ref{r54}).

\subsection{$H\left( \gamma \right) $ and $H\left( \delta |d\right) $\label%
{hgama}}

The formula (\ref{r54}) shows that all the antifields 
\begin{equation}
\chi ^{*\Delta }=\left( t^{*\mu \nu |\alpha \beta },\eta ^{*\mu \nu |\alpha
},C^{*\mu \nu }\right)  \label{r73ab}
\end{equation}
belong (non-trivially) to $H^{0}\left( \gamma \right) $. From the definition
(\ref{r50}) and recalling the general discussion from Section \ref{2} on the
relationship between the model under investigation and the 3-differential
complex, we infer that the most general $\gamma $-closed (and obviously
non-trivial) elements constructed in terms of the original tensor field are
the components of the curvature tensor (\ref{curv}) and their spacetime
derivatives, so all these pertain to $H^{0}\left( \gamma \right) $.

Using the first definition in (\ref{r51}), we notice that there is no $%
\gamma $-closed linear combination of the undifferentiated ghosts of pure
ghost number one. On behalf of the same definition, we investigate the
existence of $\gamma $-closed linear combinations in the first-order
derivatives of these ghosts. By direct computation, it is easy to see that
the most general $\gamma $-closed quantities in the first-order derivatives
of the pure ghost number one ghosts have the mixed symmetry of the tensor
field $t_{\mu \nu |\alpha \beta }$ itself 
\begin{equation}
M_{\mu \nu |\alpha \beta }=\partial _{\mu }\eta _{\alpha \beta |\nu
}-\partial _{\nu }\eta _{\alpha \beta |\mu }+\partial _{\alpha }\eta _{\mu
\nu |\beta }-\partial _{\beta }\eta _{\mu \nu |\alpha }.  \label{r75a}
\end{equation}
Howevr, with the help of the formula (\ref{r50}) it is obvious that $M_{\mu
\nu |\alpha \beta }$ is $\gamma $-exact, $M_{\mu \nu |\alpha \beta }=\gamma
t_{\mu \nu |\alpha \beta }$, and thus it must be discarded from $H^{1}\left(
\gamma \right) $ as being trivial. Along the same line, one can prove that
the only $\gamma $-closed combinations with $N\geq 2$ spacetime derivatives
of the ghosts $\eta _{\mu \nu |\alpha }$ are actually polynomials with $%
\left( N-1\right) $ derivatives in the elements $M_{\mu \nu |\alpha \beta }$%
, so they are $\gamma $-exact, and hence trivial in $H^{1}\left( \gamma
\right) $. In conclusion, there is no non-trivial object constructed out of
the ghosts $\eta _{\mu \nu |\alpha }$ and their derivatives in $H^{1}\left(
\gamma \right) $, which implies that $H^{1}\left( \gamma \right) =0$ as
there are no other ghosts of pure ghost number equal to one in the BRST
complex. The BRST complex for the model under consideration contains no
other ghosts with odd pure ghost numbers, so we conclude that 
\begin{equation}
H^{2l+1}\left( \gamma \right) =0,\;\mathrm{for}\;\mathrm{all}\;l\geq 0.
\label{hgzero}
\end{equation}

The definitions (\ref{r51}) show that the undifferentiated ghosts of pure
ghost number equal to two, $C_{\mu \nu }$, belong to $H\left( \gamma \right) 
$. The $\gamma $- closedness of $C_{\mu \nu }$ further implies that all
their derivatives are also $\gamma $-closed. Let us see which of these
derivatives are trivial. Regarding their first-order derivatives, from the
first relation in (\ref{r51}) we observe that their symmetric part is $%
\gamma $-exact 
\begin{equation}
\partial _{\left( \mu \right. }C_{\left. \nu \right) \alpha }\equiv \gamma
\left( -\frac{1}{3}\eta _{\alpha \left( \mu |\nu \right) }\right) ,
\label{r77}
\end{equation}
where $\left( \mu \nu \cdots \right) $ denotes plain symmetrization with
respect to the indices between brackets without normalization factors, such
that $\partial _{\left( \mu \right. }C_{\left. \nu \right) \alpha }$ will be
removed from $H\left( \gamma \right) $. Meanwhile, their antisymmetric part $%
\partial _{\left[ \mu \right. }C_{\left. \nu \right] \alpha }$ is not $%
\gamma $-exact, and hence can be taken as a non-trivial representative of $%
H\left( \gamma \right) $. After some calculations, we find that all the
second-order derivatives of the ghosts for ghosts are $\gamma $-exact 
\begin{equation}
\partial _{\alpha }\partial _{\beta }C_{\mu \nu }=\frac{1}{12}\gamma \left(
3\left( \partial _{\alpha }\eta _{\mu \nu |\beta }+\partial _{\beta }\eta
_{\mu \nu |\alpha }\right) +\partial _{\left[ \mu \right. }\eta _{\left. \nu
\right] \,\left( \alpha |\beta \right) }\right) ,  \label{r78}
\end{equation}
and so will be their higher-order derivatives. In conclusion, the only
non-trivial combinations in $H\left( \gamma \right) $ constructed from the
ghosts of pure ghost number equal to two are polynomials in $C_{\mu \nu }$
and $\partial _{\left[ \mu \right. }C_{\left. \nu \right] \alpha }$.
Combining this result with the previous one on $H^{0}\left( \gamma \right) $
being non-vanishing, we have actually proved that only the even
cohomological spaces of the exterior longitudinal derivative, $H^{2l}\left(
\gamma \right) $ with $l\geq 0$, are non-vanishing.

Under these circumstances, it follows that the equation (\ref{r67})
possesses non-trivial solutions only for $I=2J$, where the general form of $%
a_{2J}$ for $J>0$ is (up to irrelevant, $\gamma $-exact contributions) 
\begin{equation}
a_{I}\equiv a_{2J}=\alpha _{2J}\left( \left[ \chi ^{*\Delta }\right] ,\left[
F_{\mu \nu \lambda |\alpha \beta \gamma }\right] \right) e^{2J}\left( C_{\mu
\nu },\partial _{\left[ \mu \right. }C_{\left. \nu \right] \alpha }\right)
,\;J>0,  \label{r79}
\end{equation}
where the notation $f\left( \left[ q\right] \right) $ means that $f$ depends
on $q$ and its spacetime derivatives up to a finite order. The coefficients $%
\alpha _{2J}$ are $\gamma $-invariant 
\begin{equation}
\gamma \alpha _{2J}=0,  \label{r79a}
\end{equation}
and exhibit the properties $\varepsilon \left( \alpha _{2J}\right) =0$, $%
\mathrm{pgh}\left( \alpha _{2J}\right) =0$ and $\mathrm{agh}\left( \alpha
_{2J}\right) =2J$, while the symbol $e^{2J}$ stands for a generic notation
of the elements with pure ghost number equal to $2J$ of a basis in the space
of polynomials in $C_{\mu \nu }$ and $\partial _{\left[ \mu \right.
}C_{\left. \nu \right] \alpha }$. The objects $\alpha _{2J}$ (obviously
non-trivial in $H^{0}\left( \gamma \right) $) were taken to have a bounded
number of derivatives, and therefore they are polynomials in the antifields $%
\chi ^{*\Delta }$, in the curvature tensor $F_{\mu \nu \lambda |\alpha \beta
\gamma }$, as well as in their derivatives. Due to their $\gamma $%
-closedness, they are called invariant polynomials. At zero antighost
number, the invariant polynomials are polynomials in the curvature tensor $%
F_{\mu \nu \lambda |\alpha \beta \gamma }$ and its derivatives. The result
that we can replace the equation (\ref{r64}) with the less obvious one (\ref
{r67}) is a nice consequence of the fact that the cohomology of the exterior
spacetime differential is trivial in the space of invariant polynomials at
strictly positive antighost numbers. This means that if the invariant
polynomial $\alpha _{I}$ of strictly positive antighost number is
annihilated by $d$, then it can be written like the $d$-variation of
precisely an invariant polynomial. For details, see Section 3 from \cite
{0402099}.

Replacing the solution (\ref{r79}) in the equation (\ref{r65}) for $I=2J$
and taking into account the definitions (\ref{r51}), we remark that a
necessary (but not sufficient) condition for the existence of (non-trivial)
solutions $a_{2J-1}$ is that the invariant polynomials $\alpha _{2J}$ from (%
\ref{r79}) are (non-trivial) objects from the local cohomology of the
Koszul-Tate differential $H\left( \delta |d\right) $ at antighost number $%
2J>0$ and pure ghost number equal to zero\footnote{%
We recall that the local cohomology $H\left( \delta |d\right) $ is
completely trivial at both strictly positive antighost \textit{and} pure
ghost numbers (for instance, see~\cite{gen1}, Theorem 5.1 and~\cite{commun1}%
).}, $\alpha _{2J}\in H_{2J}\left( \delta |d\right) $, i.e. 
\begin{equation}
\delta \alpha _{2J}=\partial _{\mu }j^{\mu },\;\varepsilon \left( j^{\mu
}\right) =1,\;\mathrm{agh}\left( j^{\mu }\right) =2J-1,\;\mathrm{pgh}\left(
j^{\mu }\right) =0.  \label{r80}
\end{equation}
Consequently, we need to investigate some of the main properties of the
local cohomology of the Koszul-Tate differential at pure ghost number zero
and strictly positive antighost numbers in order to completely determine the
component $a_{2J}$ of highest antighost number in the first-order
deformation. As we have discussed in Section \ref{2}, the free model under
study is a normal gauge theory of Cauchy order equal to three. Using the
general results from~\cite{gen1} (also see~\cite{lingr} and~\cite{multi,gen2}%
), one can state that the local cohomology of the Koszul-Tate differential
at pure ghost number zero is trivial at antighost numbers strictly greater
than its Cauchy order 
\begin{equation}
H_{k}\left( \delta |d\right) =0,\;k>3.  \label{r81}
\end{equation}
Moreover, if the invariant polynomial $\alpha _{k}$, with $\mathrm{agh}%
\left( \alpha _{k}\right) =k\geq 3$, is trivial in $H_{k}\left( \delta
|d\right) $, then it can be taken to be trivial also in $H_{k}^{\mathrm{inv}%
}\left( \delta |d\right) $%
\begin{equation}
\left( \alpha _{k}=\delta b_{k+1}+\partial _{\mu }\stackrel{(k)}{c}^{\mu },\;%
\mathrm{agh}\left( \alpha _{k}\right) =k\geq 3\right) \Rightarrow \alpha
_{k}=\delta \beta _{k+1}+\partial _{\mu }\stackrel{(k)}{\gamma }^{\mu },
\label{r81d}
\end{equation}
where $\beta _{k+1}$ and $\stackrel{(k)}{\gamma }^{\mu }$ are invariant
polynomials. [An element of $H_{k}^{\mathrm{inv}}\left( \delta |d\right) $
is defined via an equation similar to (\ref{r80}) for $2J\rightarrow k$, but
with the corresponding current an invariant polynomial.] The result (\ref
{r81d}) is proved in Theorem 4.1 from \cite{0402099}. It is important since
together with (\ref{r81}) ensure that all the local cohomology of the
Koszul-Tate differential in the space of invariant polynomials is trivial in
antighost numbers strictly greater than three 
\begin{equation}
H_{k}^{\mathrm{inv}}\left( \delta |d\right) =0,\;k>3.  \label{r81c}
\end{equation}
Using the definitions (\ref{r53}), we can organize the non-trivial
representatives of $\left( H_{k}\left( \delta |d\right) \right) _{k\geq 2}$
(at pure ghost number equal to zero) and $\left( H_{k}^{\mathrm{inv}}\left(
\delta |d\right) \right) _{k\geq 2}$ like 
\begin{equation}
\begin{array}{cc}
\mathrm{agh} & 
\begin{array}{l}
\mathrm{non-trivial\;representatives} \\ 
\mathrm{spanning\;}H_{k}\left( \delta |d\right) \;\mathrm{and}\;H_{k}^{%
\mathrm{inv}}\left( \delta |d\right)
\end{array}
\\ 
k>3 & \mathrm{none} \\ 
k=3 & C^{*\mu \nu } \\ 
k=2 & \eta ^{*\mu \nu |\alpha }
\end{array}
.  \label{tabledelta}
\end{equation}
With the help of the above representatives we can construct in principle
other non-trivial elements from $H\left( \delta |d\right) $ and $H^{\mathrm{%
inv}}\left( \delta |d\right) $ at strictly positive antighost numbers, which
explicitly depend on the spacetime co-ordinates. For instance, the object $%
\eta _{\mu \nu |\alpha }^{*}f^{\mu \nu }x^{\alpha }$, with $f^{\mu \nu }$
some antisymmetric constants, belongs to both $H_{2}\left( \delta |d\right) $
and $H_{2}^{\mathrm{inv}}\left( \delta |d\right) $. However, we will discard
such elements during the deformation procedure since they would break the
Poincar\'{e} invariance of the interactions. In contrast to the groups $%
\left( H_{k}\left( \delta |d\right) \right) _{k\geq 2}$ and $\left( H_{k}^{%
\mathrm{inv}}\left( \delta |d\right) \right) _{k\geq 2}$, which are
finite-dimensional, the cohomology $H_{1}\left( \delta |d\right) $ at pure
ghost number zero, that is related to global symmetries and ordinary
conservation laws, is infinite-dimensional since the theory is free.
Fortunately, it will not be needed in the sequel.

The above results on $H\left( \delta |d\right) $ and $H^{\mathrm{inv}}\left(
\delta |d\right) $ in strictly positive antighost number are important
because they control the obstructions to removing the antifields from the
first-order deformation. Indeed, due to (\ref{r81c}) and (\ref{hgzero}) it
follows that we can successively eliminate all the pieces of antighost
number strictly greater than two from the non-integrated density of the
first-order deformation by adding only trivial terms (for details, see
Section 5 from \cite{0402099}), so we can take, without loss of non-trivial
objects, the condition 
\begin{equation}
0\leq I=2J\leq 2  \label{r81b}
\end{equation}
in the development (\ref{r63}), which leaves us with a single eligible,
strictly positive value, $I=2J=2$.

\subsection{The case $I=2$\label{5.2}}

Thus, for $I=2J=2$ we finally obtain that the expansion (\ref{r63}) becomes 
\begin{equation}
a=a_{0}+a_{1}+a_{2},  \label{r81a}
\end{equation}
where its last component is written (up to $\gamma $-exact objects) in the
form 
\begin{equation}
a_{2}=\alpha _{2}\left( \left[ t^{*\mu \nu |\alpha \beta }\right] ,\left[
\eta ^{*\mu \nu |\alpha }\right] ,\left[ F_{\mu \nu \lambda |\alpha \beta
\gamma }\right] \right) e^{2}\left( C_{\mu \nu },\partial _{\left[ \mu
\right. }C_{\left. \nu \right] \alpha }\right) ,  \label{r82}
\end{equation}
with the elements of pure ghost number two spanned by 
\begin{equation}
\left( C_{\mu \nu },\partial _{\left[ \mu \right. }C_{\left. \nu \right]
\alpha }\right) .  \label{r83}
\end{equation}
Taking into account the result from (\ref{tabledelta}) at $k=2$, we get that 
\begin{equation}
a_{2}=\eta _{\mu \nu |\alpha }^{*}\left( f^{\mu \nu \alpha \beta \gamma
}C_{\beta \gamma }+\bar{f}^{\mu \nu \alpha \beta \gamma \lambda }\partial
_{\left[ \beta \right. }C_{\left. \gamma \right] \lambda }\right) ,
\label{r88}
\end{equation}
where $f^{\mu \nu \alpha \beta \gamma }$ and $\bar{f}^{\mu \nu \alpha \beta
\gamma \lambda }$ must be non-derivative constants. In the meantime, $f^{\mu
\nu \alpha \beta \gamma }$ and $\bar{f}^{\mu \nu \alpha \beta \gamma \lambda
}$ cannot be antisymmetric in all indices $\left\{ \mu ,\nu ,\alpha \right\} 
$ (because in this event the identity $\eta ^{*\left[ \mu \nu |\alpha
\right] }\equiv 0$ maps the corresponding terms to zero), which eventually
leaves one candidate for $a_{2}$ 
\begin{equation}
a_{2}=c\eta ^{*\mu \nu |\alpha }\partial _{\left[ \mu \right. }C_{\left. \nu
\right] \alpha },  \label{r89}
\end{equation}
with $c$ an arbitrary real constant. However, this term is easily shown to
be trivial ($\gamma $-exact) on account of the first definition in (\ref{r51}%
) and of the identity $\eta ^{*\left[ \mu \nu |\alpha \right] }\equiv 0$,
which allows us to add to $a_{2}$ any quantity proportional with $\eta
^{*\mu \nu |\alpha }\partial _{\left[ \mu \right. }C_{\left. \nu \alpha
\right] }$ since it vanishes identically 
\begin{equation}
c\eta ^{*\mu \nu |\alpha }\partial _{\left[ \mu \right. }C_{\left. \nu
\right] \alpha }=c\eta ^{*\mu \nu |\alpha }\left( \partial _{\left[ \mu
\right. }C_{\left. \nu \right] \alpha }-\frac{2}{3}\partial _{\left[ \alpha
\right. }C_{\left. \mu \nu \right] }\right) =\gamma \left( -\frac{c}{3}\eta
^{*\mu \nu |\alpha }\eta _{\mu \nu |\alpha }\right) ,  \label{r90}
\end{equation}
and so it can be discarded from (\ref{r89}) by setting 
\begin{equation}
c=0.  \label{r91}
\end{equation}
So far we have shown that there is no non-trivial $a_{2}$ in the right-hand
side of (\ref{r81a}) 
\begin{equation}
a_{2}=0.  \label{r93a}
\end{equation}
It is worth noticing that at this stage we have not used any a priori
restriction on the number of derivatives from $a_{2}$, except that it is
finite, but only the general requirements of smooth, local,
Lorentz-covariant and Poincar\'{e}-invariant deformations. The assumption
that the interactions contain at most two derivatives will only be needed
below.

\subsection{The case $I=0$\label{5.3}}

Consequently, we pass to the next value of the maximum antighost number in
the expansion (\ref{r63}), which, according to the restriction (\ref{r81b}),
excludes the value $I=1$. Thus, we are only left with the possibility that
the non-integrated density of the first-order deformation reduces to its
antighost number zero component, which is nothing but the deformed
Lagrangian at the first order in the coupling constant 
\begin{equation}
a=a_{0}\left( \left[ t_{\mu \nu |\alpha \beta }\right] \right) ,  \label{r94}
\end{equation}
which must obey the equation 
\begin{equation}
\gamma a_{0}=\partial _{\mu }m^{\mu }.  \label{r95}
\end{equation}
There are two main types of solutions to the last equation. The first one
corresponds to $m^{\mu }=0$ and is given by functions in the field $t_{\mu
\nu |\alpha \beta }$ and its derivatives that are invariant under the gauge
transformations (\ref{r8}). As the components of the curvature tensor are
the most general gauge invariant objects, it follows that 
\begin{equation}
\gamma a_{0}^{\prime }=0\Rightarrow a_{0}^{\prime }=a_{0}^{\prime }\left(
\left[ F_{\mu \nu \lambda |\alpha \beta \gamma }\right] \right) .
\label{r96}
\end{equation}
At this point we ask that the deformed gauge theory preserves the Cauchy
order of the uncoupled model, which enforces the requirement that the
interacting Lagrangian is of maximum order equal to two in the spacetime
derivatives of the tensor field $t_{\mu \nu |\alpha \beta }$ at each order
in the coupling constant. In turn, this requirement leads to $a_{0}^{\prime
}=0$ (we have excluded the solutions linear in $\left[ F_{\mu \nu \lambda
|\alpha \beta \gamma }\right] $, as they obviously reduce to total
divergences, and thus give a vanishing $S_{1}$).

The second type of solutions is associated with $m^{\mu }\neq 0$, being
understood that we maintain the restriction on the derivative order of $%
a_{0} $ and discard the divergence-like solutions $a_{0}=\partial _{\mu
}u^{\mu }$. Denoting the Euler-Lagrange derivatives of $a_{0}$ by $A^{\mu
\nu |\alpha \beta }\equiv \delta a_{0}/\delta t_{\mu \nu |\alpha \beta }$
and using the formula (\ref{r50}), the equation (\ref{r95}) implies that 
\begin{equation}
\partial _{\mu }A^{\mu \nu |\alpha \beta }=0,  \label{r97}
\end{equation}
where the tensor $A^{\mu \nu |\alpha \beta }$ is imposed to contain at most
two derivatives, to have the mixed symmetry of $t_{\mu \nu |\alpha \beta }$
and to fulfill the Bianchi I identity $A^{\left[ \mu \nu |\alpha \right]
\beta }\equiv 0$.

According to the discussion from the end of Section \ref{2} (see the
formulas (\ref{r40a}--\ref{r40b})), the general solution to (\ref{r97}) is 
\begin{equation}
\frac{\delta a_{0}}{\delta t_{\mu \nu |\alpha \beta }}\equiv A^{\mu \nu
|\alpha \beta }=\partial _{\rho }\partial _{\gamma }\tilde{\Phi}^{\mu \nu
\rho |\alpha \beta \gamma }+c\left( \sigma ^{\mu \alpha }\sigma ^{\nu \beta
}-\sigma ^{\mu \beta }\sigma ^{\nu \alpha }\right) ,  \label{r97b}
\end{equation}
where $\tilde{\Phi}^{\mu \nu \rho |\alpha \beta \gamma }$ has the mixed
symmetry of the curvature tensor. The second term in (\ref{r97b}) is
non-trivial and generates a cosmological-like term 
\begin{equation}
a_{0}^{(1)}=2ct,  \label{r97bb}
\end{equation}
where $t$ is the double trace of the tensor field $t_{\mu \nu |\alpha \beta
} $. It verifies the equation 
\begin{equation}
\gamma a_{0}^{(1)}=\partial _{\mu }m^{(1)\mu },\;m^{(1)\mu }=8c\eta
_{\;\;\;\;\;\;\alpha }^{\mu \alpha |},  \label{r97bc}
\end{equation}
so we can write that 
\begin{equation}
a_{0}=a_{0}^{(1)}+a_{0}^{(2)},  \label{r97bd}
\end{equation}
with 
\begin{equation}
\gamma a_{0}^{(2)}=\partial _{\mu }m^{(2)\mu }  \label{r97be}
\end{equation}
and 
\begin{equation}
\frac{\delta a_{0}^{(2)}}{\delta t_{\mu \nu |\alpha \beta }}=\partial _{\rho
}\partial _{\gamma }\tilde{\Phi}^{\mu \nu \rho |\alpha \beta \gamma }.
\label{r97bf}
\end{equation}
In the sequel we investigate the form of $a_{0}^{(2)}$. Imposing that $%
A^{\mu \nu |\alpha \beta }$ contains at most two derivatives, we find that $%
\tilde{\Phi}^{\mu \nu \rho |\alpha \beta \gamma }$ involves only the
undifferentiated tensor field $t_{\mu \nu |\alpha \beta }$. Let $N$ be a
derivation in the algebra of the fields $t_{\mu \nu |\alpha \beta }$ and of
their derivatives that counts the powers of the fields and their
derivatives, defined by 
\begin{equation}
N=\sum\limits_{n\geq 0}\left( \partial _{\mu _{1}}\cdots \partial _{\mu
_{n}}t_{\mu \nu |\alpha \beta }\right) \frac{\partial }{\partial \left(
\partial _{\mu _{1}}\cdots \partial _{\mu _{n}}t_{\mu \nu |\alpha \beta
}\right) }.  \label{r97c}
\end{equation}
Then, it is easy to see that for every non-integrated density $u$, we have
that 
\begin{equation}
Nu=t_{\mu \nu |\alpha \beta }\frac{\delta u}{\delta t_{\mu \nu |\alpha \beta
}}+\partial _{\mu }s^{\mu },  \label{r97d}
\end{equation}
where $\delta u/\delta t_{\mu \nu |\alpha \beta }$ denotes the variational
derivative of $u$. If $u$ is a homogeneous polynomial of order $p>0$ in the
fields and their derivatives, then $Nu=pu$, such that 
\begin{equation}
u=\frac{1}{p}t_{\mu \nu |\alpha \beta }\frac{\delta u}{\delta t_{\mu \nu
|\alpha \beta }}+\partial _{\mu }\left( \frac{1}{p}s^{\mu }\right) .
\label{r97e}
\end{equation}
As $a_{0}^{(2)}$ can always be decomposed as a sum of homogeneous
polynomials of various orders in the fields and their derivatives, it is
enough to analyze the equation (\ref{r97be}) for a fixed value of $p$.
Setting $u=a_{0}^{(2)}$ in (\ref{r97e}) and using (\ref{r97bf}), we find
that 
\begin{equation}
a_{0}^{(2)}=\frac{1}{p}t_{\mu \nu |\alpha \beta }\partial _{\rho }\partial
_{\gamma }\tilde{\Phi}^{\mu \nu \rho |\alpha \beta \gamma }+\partial _{\mu }%
\tilde{s}^{\mu }.  \label{r97f}
\end{equation}
Integrating twice by parts in (\ref{r97f}) and taking into account the mixed
symmetry of $\tilde{\Phi}^{\mu \nu \rho |\alpha \beta \gamma }$, we infer
that 
\begin{equation}
a_{0}^{(2)}=kF_{\mu \nu \rho |\alpha \beta \gamma }\tilde{\Phi}^{\mu \nu
\rho |\alpha \beta \gamma }+\partial _{\mu }l^{\mu },  \label{r97g}
\end{equation}
with $k=1/9p$. Acting now with $\gamma $ on (\ref{r97g}), we obtain that 
\begin{equation}
\gamma a_{0}^{(2)}=-4k\eta _{\xi \eta |\varepsilon }\partial _{\delta
}\left( F_{\mu \nu \rho |\alpha \beta \gamma }\frac{\partial \tilde{\Phi}%
^{\mu \nu \rho |\alpha \beta \gamma }}{\partial t_{\delta \varepsilon |\xi
\eta }}\right) +\partial _{\mu }\bar{l}^{\mu },  \label{r97h}
\end{equation}
for some $\bar{l}^{\mu }$. From (\ref{r97h}) we observe that $a_{0}^{(2)}$
satisfies the equation (\ref{r97be}) if and only if 
\begin{equation}
\partial _{\delta }\left( F_{\mu \nu \rho |\alpha \beta \gamma }\frac{%
\partial \tilde{\Phi}^{\mu \nu \rho |\alpha \beta \gamma }}{\partial
t_{\delta \varepsilon |\xi \eta }}\right) =0.  \label{r97i}
\end{equation}
Since the quantity between parentheses in (\ref{r97i}) has the same mixed
symmetry like the tensor field $t_{\delta \varepsilon |\xi \eta }$, with the
help of the relations (\ref{r40a}--\ref{r40b}) we determine that 
\begin{equation}
F_{\mu \nu \rho |\alpha \beta \gamma }\frac{\partial \tilde{\Phi}^{\mu \nu
\rho |\alpha \beta \gamma }}{\partial t_{\delta \varepsilon |\xi \eta }}%
=\partial _{\varphi }\partial _{\theta }\psi ^{\delta \varepsilon \varphi
|\xi \eta \theta },  \label{r97j}
\end{equation}
for some $\psi ^{\delta \varepsilon \varphi |\xi \eta \theta }$ with the
mixed symmetry of the curvature tensor, which depends only on the
undifferentiated tensor field $t_{\mu \nu |\alpha \beta }$. Computing the
left-hand side of (\ref{r97j}), we arrive at 
\begin{eqnarray}
&&F_{\mu \nu \rho |\alpha \beta \gamma }\frac{\partial \tilde{\Phi}^{\mu \nu
\rho |\alpha \beta \gamma }}{\partial t_{\delta \varepsilon |\xi \eta }}%
=\partial _{\varphi }\partial _{\theta }\left( 9t_{\mu \nu |\alpha \beta }%
\frac{\partial \tilde{\Phi}^{\mu \nu \varphi |\alpha \beta \theta }}{%
\partial t_{\delta \varepsilon |\xi \eta }}\right)  \nonumber \\
&&-9\frac{\partial ^{2}\tilde{\Phi}^{\mu \nu \rho |\alpha \beta \gamma }}{%
\partial t_{\delta \varepsilon |\xi \eta }\partial t_{\delta ^{\prime
}\varepsilon ^{\prime }|\xi ^{\prime }\eta ^{\prime }}}\left( 2\left(
\partial _{\rho }t_{\mu \nu |\alpha \beta }\right) \left( \partial _{\gamma
}t_{\delta ^{\prime }\varepsilon ^{\prime }|\xi ^{\prime }\eta ^{\prime
}}\right) +t_{\mu \nu |\alpha \beta }\partial _{\rho }\partial _{\gamma
}t_{\delta ^{\prime }\varepsilon ^{\prime }|\xi ^{\prime }\eta ^{\prime
}}\right)  \nonumber \\
&&-9\frac{\partial ^{3}\tilde{\Phi}^{\mu \nu \rho |\alpha \beta \gamma }}{%
\partial t_{\delta \varepsilon |\xi \eta }\partial t_{\delta ^{\prime
}\varepsilon ^{\prime }|\xi ^{\prime }\eta ^{\prime }}\partial t_{\delta
^{\prime \prime }\varepsilon ^{\prime \prime }|\xi ^{\prime \prime }\eta
^{\prime \prime }}}t_{\mu \nu |\alpha \beta }\left( \partial _{\rho
}t_{\delta ^{\prime }\varepsilon ^{\prime }|\xi ^{\prime }\eta ^{\prime
}}\right) \left( \partial _{\gamma }t_{\delta ^{\prime \prime }\varepsilon
^{\prime \prime }|\xi ^{\prime \prime }\eta ^{\prime \prime }}\right) .
\label{r97k}
\end{eqnarray}
The right-hand side of (\ref{r97k}) can be written in the form of the
right-hand side from (\ref{r97j}) if and only if 
\begin{equation}
9t_{\mu \nu |\alpha \beta }\frac{\partial \tilde{\Phi}^{\mu \nu \varphi
|\alpha \beta \theta }}{\partial t_{\delta \varepsilon |\xi \eta }}=\psi
^{\delta \varepsilon \varphi |\xi \eta \theta },  \label{r97l}
\end{equation}
\begin{equation}
\frac{\partial ^{2}\tilde{\Phi}^{\mu \nu \rho |\alpha \beta \gamma }}{%
\partial t_{\delta \varepsilon |\xi \eta }\partial t_{\delta ^{\prime
}\varepsilon ^{\prime }|\xi ^{\prime }\eta ^{\prime }}}=0,\;\frac{\partial
^{3}\tilde{\Phi}^{\mu \nu \rho |\alpha \beta \gamma }}{\partial t_{\delta
\varepsilon |\xi \eta }\partial t_{\delta ^{\prime }\varepsilon ^{\prime
}|\xi ^{\prime }\eta ^{\prime }}\partial t_{\delta ^{\prime \prime
}\varepsilon ^{\prime \prime }|\xi ^{\prime \prime }\eta ^{\prime \prime }}}%
=0.  \label{r97m}
\end{equation}
On the one hand, the requirements (\ref{r97m}) restrict $\tilde{\Phi}^{\mu
\nu \rho |\alpha \beta \gamma }$ to be linear in $t_{\mu \nu |\alpha \beta }$
and, on the other hand, we have the condition that $\tilde{\Phi}^{\mu \nu
\rho |\alpha \beta \gamma }$ has the same mixed symmetry like the curvature
tensor. These considerations fix $\tilde{\Phi}^{\mu \nu \rho |\alpha \beta
\gamma }$ to be precisely of the type 
\begin{equation}
\tilde{\Phi}^{\mu \nu \rho |\alpha \beta \gamma }=k^{\prime }\Phi ^{\mu \nu
\rho |\alpha \beta \gamma },  \label{r97n}
\end{equation}
where $k^{\prime }$ is a real constant and $\Phi ^{\mu \nu \rho |\alpha
\beta \gamma }$ is the tensor (\ref{r40d}) involved in the functions (\ref
{r40c}) that yield the free field equations. Meanwhile, (\ref{r97n}) fixes
the value of $p$ from (\ref{r97f}) to $p=2$. By direct computation we deduce
that (\ref{r97l}) is also satisfied and get that 
\begin{equation}
\psi ^{\delta \varepsilon \varphi |\xi \eta \theta }=9k^{\prime }\Phi
^{\delta \varepsilon \varphi |\xi \eta \theta }.  \label{r97o}
\end{equation}
Inserting (\ref{r97n}) in (\ref{r97f}) for $p=2$, on behalf of (\ref{r40c})
we infer that 
\begin{equation}
a_{0}^{(2)}=k^{\prime }t_{\mu \nu |\alpha \beta }T^{\mu \nu |\alpha \beta
}+\partial _{\mu }l^{\mu },  \label{r97p}
\end{equation}
and hence (\ref{r97p}) is (up to an irrelevant divergence) proportional with
the original Lagrangian. This solution is however trivial in $H^{0}\left(
s|d\right) $ since it can be expressed like 
\begin{equation}
a_{0}^{(2)}=sb+\partial _{\mu }v^{\mu },\;\mathrm{gh}\left( b\right) =-1,\;%
\mathrm{gh}\left( v^{\mu }\right) =0,  \label{r97q}
\end{equation}
where 
\begin{eqnarray}
b &=&4k^{\prime }\left( t^{*\mu \nu |\alpha \beta }t_{\mu \nu |\alpha \beta
}+\eta ^{*\mu \nu |\alpha }\eta _{\mu \nu |\alpha }+C^{*\mu \nu }C_{\mu \nu
}\right) ,  \label{r97r} \\
v^{\mu } &=&\left( l^{\mu }-16k^{\prime }t^{*\mu \nu |\alpha \beta }\eta
_{\alpha \beta |\nu }-12k^{\prime }\eta ^{*\alpha \beta |\mu }C_{\alpha
\beta }\right) .  \label{r97s}
\end{eqnarray}
Then, in agreement with the discussion from the beginning of this section,
the solution (\ref{r97q}) can be safely removed from the first-order
deformation by replacing it with 
\begin{equation}
a_{0}^{(2)}=0.  \label{r97t}
\end{equation}
From (\ref{r97bb}) for $c=1/2$, using (\ref{r97bd}), and relying on the
results contained in the previous subsections, we conclude that 
\begin{equation}
S_{1}=\int d^{D}x\,t  \label{r98}
\end{equation}
represents the only non-trivial first-order deformation of the solution to
the master equation for the tensor $t_{\mu \nu |\alpha \beta }$. Moreover,
it is consistent to all orders in the coupling constant. Indeed, as $\left(
S_{1},S_{1}\right) =0$, the equation (\ref{r61}) that describes the
second-order deformation is satisfied with the choice 
\begin{equation}
S_{2}=0,  \label{r99}
\end{equation}
while the remaining higher-order equations are fulfilled for 
\begin{equation}
S_{3}=S_{4}=\cdots =0,  \label{r100}
\end{equation}
and hence there are no non-trivial self-interactions for the tensor field $%
t_{\mu \nu |\alpha \beta }$.

The main conclusion of this section is that, under the general conditions of
smoothness, locality, Lorentz covariance and Poincar\'{e} invariance of the
deformations, combined with the requirement that the interacting Lagrangian
is at most second-order derivative, there are no consistent, non-trivial
self-interactions for the massless tensor field with the mixed symmetry of
the Riemann tensor. The only piece that can be added to the original
Lagrangian is a cosmological-like term, which does not modify the original
gauge transformations.

\section{Interactions with the Pauli-Fierz theory\label{6}}

We have argued in the previous section that there are no consistent
self-interactions that can be added to the free action of the massless
tensor field $t_{\mu \nu |\alpha \beta }$. In the sequel we investigate if
there exist local, smooth, Lorentz-covariant and Poincar\'{e}-invariant,
consistent interactions between such a tensor field and a non-interacting
massless spin-2 field $h_{\mu \nu }$, described by the Pauli-Fierz action~%
\cite{pf}. We maintain the restriction on the maximum derivative order of
the interactions being equal to two. The self-interactions of a single
massless spin-2 field have been extensively studied in the literature and
are known to lead to the Einstein-Hilbert action with a cosmological term.
We will mainly focus on the cross-couplings, i.e. on the interactions that
mix the fields $t_{\mu \nu |\alpha \beta }$ and $h_{\mu \nu }$, and will not
insist on the cohomological construction of the Einstein-Hilbert action with
a cosmological term, which can be found in detail in~\cite{multi}.

\subsection{Free model and accompanying BRST symmetry\label{6.1}}

We start from a free action, written as the sum between (\ref{r1}) and the
Pauli-Fierz action in $D\geq 5$ spacetime dimensions 
\begin{equation}
S_{0}\left[ t_{\mu \nu |\alpha \beta },h_{\mu \nu }\right] =S_{0}\left[
t_{\mu \nu |\alpha \beta }\right] +S_{0}^{\mathrm{PF}}\left[ h_{\mu \nu
}\right] ,  \label{r101}
\end{equation}
with 
\begin{eqnarray}
S_{0}^{\mathrm{PF}}\left[ h_{\mu \nu }\right] &=&\int d^{D}x\left( -\frac{1}{%
2}\left( \partial ^{\rho }h^{\mu \nu }\right) \left( \partial _{\rho }h_{\mu
\nu }\right) +\left( \partial _{\rho }h^{\rho \mu }\right) \left( \partial
^{\lambda }h_{\lambda \mu }\right) \right.  \nonumber \\
&&\left. -\left( \partial ^{\rho }h\right) \left( \partial ^{\lambda
}h_{\lambda \rho }\right) +\frac{1}{2}\left( \partial ^{\rho }h\right)
\left( \partial _{\rho }h\right) \right) ,  \label{r102}
\end{eqnarray}
where $h_{\mu \nu }$ is symmetric and $h$ denotes its trace. Action (\ref
{r102}) is invariant under the abelian and irreducible gauge transformations 
\begin{equation}
\delta _{\epsilon }h_{\mu \nu }=\partial _{\left( \mu \right. }\epsilon
_{\left. \nu \right) }.  \label{r103}
\end{equation}
The presence of the gauge transformations (\ref{r103}) shows that the
functions that define the field equations of the Pauli-Fierz action 
\begin{equation}
\frac{\delta S_{0}^{\mathrm{PF}}}{\delta h^{\mu \nu }}\equiv -2H_{\mu \nu
}\approx 0  \label{r104}
\end{equation}
are not all independent, but satisfy the Noether identities 
\begin{equation}
\partial ^{\mu }H_{\mu \nu }=0.  \label{r105}
\end{equation}
In the above, $H_{\mu \nu }$ represents the linearized Einstein tensor 
\begin{equation}
H_{\mu \nu }=K_{\mu \nu }-\frac{1}{2}\sigma _{\mu \nu }K,\;H_{\mu \nu
}=H_{\nu \mu },  \label{r106}
\end{equation}
with $K_{\mu \nu }$ the linearized Ricci tensor and $K$ the linearized
scalar curvature, which are defined with the help of the linearized Riemann
tensor 
\begin{equation}
K_{\mu \nu |\alpha \beta }=-\frac{1}{2}\left( \partial _{\mu }\partial
_{\alpha }h_{\nu \beta }-\partial _{\nu }\partial _{\alpha }h_{\mu \beta
}-\partial _{\mu }\partial _{\beta }h_{\nu \alpha }+\partial _{\nu }\partial
_{\beta }h_{\mu \alpha }\right)  \label{r107}
\end{equation}
via its simple and respectively double trace $K_{\mu \nu }=K_{\;\;\mu
|\alpha \nu }^{\alpha }$, $K=K_{\;\;\mu }^{\mu }$. The linearized Riemann
tensor $K_{\mu \nu |\alpha \beta }$ exhibits the same symmetries and
satisfies the same identity (\ref{r5}) as the tensor field $t_{\mu \nu
|\alpha \beta }$, but in addition fulfills the Bianchi II identity 
\begin{equation}
\partial _{\left[ \lambda \right. }K_{\left. \mu \nu \right] \,|\alpha \beta
}\equiv 0.  \label{r109}
\end{equation}
The most general gauge invariant objects that can be constructed from $%
h_{\mu \nu }$ are the linearized Riemann tensor $K_{\mu \nu |\alpha \beta }$
and its spacetime derivatives. The Pauli-Fierz action alone describes a free
gauge theory of Cauchy order equal to two, so the Cauchy order of the theory
(\ref{r101}) is equal to three.

The main features of the Pauli-Fierz theory can be understood in an elegant
fashion via the generalized differential complex $\Omega _{2}\left( \mathcal{%
M}\right) $ introduced in Section \ref{2}. An interesting result refers to
the generalized cohomology of $\bar{d}$ on $\Omega _{2}\left( \mathcal{M}%
\right) $, where $\mathcal{M}$ has the trivial topology of $\mathbb{R}^{D}$,
combined with the operation of generalized Hodge duality. Let us consider a
symmetric, covariant tensor field $\bar{H}^{\mu \nu }$, subject to the
equation 
\begin{equation}
\partial _{\mu }\bar{H}^{\mu \nu }=0.  \label{r109c}
\end{equation}
Then, there exists a tensor $\bar{\Phi}^{\mu \alpha |\nu \beta }$ with the
mixed symmetry of the linearized Riemann tensor, such that 
\begin{equation}
\bar{H}^{\mu \nu }=\partial _{\alpha }\partial _{\beta }\bar{\Phi}^{\mu
\alpha |\nu \beta }+c\sigma ^{\mu \nu },  \label{r109d}
\end{equation}
with $c$ an arbitrary real constant. The above statement can be easily
verified with respect to the linearized Einstein tensor (\ref{r106}), which
satisfies the Noether identity (\ref{r105}) and can indeed be written in the
form (\ref{r109d}) for $c=0$%
\begin{equation}
H^{\mu \nu }=\partial _{\alpha }\partial _{\beta }\Phi ^{\mu \alpha |\nu
\beta },  \label{r109e}
\end{equation}
where the corresponding $\Phi ^{\mu \alpha |\nu \beta }$ reads as 
\begin{eqnarray}
\Phi ^{\mu \alpha |\nu \beta } &=&\frac{1}{2}\left( -h^{\mu \nu }\sigma
^{\alpha \beta }+h^{\alpha \nu }\sigma ^{\mu \beta }+h^{\mu \beta }\sigma
^{\alpha \nu }\right.  \nonumber \\
&&\left. -h^{\alpha \beta }\sigma ^{\mu \nu }+\left( \sigma ^{\mu \nu
}\sigma ^{\alpha \beta }-\sigma ^{\alpha \nu }\sigma ^{\mu \beta }\right)
h\right) .  \label{r109f}
\end{eqnarray}

The overall BRST complex comprises the BRST generators introduced in Section 
\ref{3} and associated with the theory (\ref{r1}), as well as the
Pauli-Fierz field $h_{\mu \nu }$, the fermionic ghost $\eta _{\mu }$
corresponding to the gauge invariances of (\ref{r102}), together with the
antifields $h^{*\mu \nu }$ and $\eta ^{*\mu }$ from the Pauli-Fierz sector.
The BRST differential of the entire free gauge theory splits like in (\ref
{r45}), where the actions of $\gamma $ and $\delta $ on the former BRST
generators are expressed by the formulas (\ref{r50}--\ref{r53}), while on
the latter ones are defined by 
\begin{equation}
\gamma h_{\mu \nu }=\partial _{\left( \mu \right. }\eta _{\left. \nu \right)
},\;\gamma \eta _{\mu }=0,  \label{r110}
\end{equation}
\begin{equation}
\gamma h^{*\mu \nu }=0=\gamma \eta ^{*\mu },  \label{r110a}
\end{equation}
\begin{equation}
\delta h_{\mu \nu }=0=\delta \eta _{\mu },  \label{r111a}
\end{equation}
\begin{equation}
\delta h^{*\mu \nu }=2H_{\mu \nu },\;\delta \eta ^{*\mu }=-2\partial _{\nu
}h^{*\nu \mu }.  \label{r111}
\end{equation}
The pure ghost number and antighost number of the BRST generators can
partially be found in (\ref{r46}--\ref{r49}), while for the Pauli-Fierz
field/ghost/antifield sector are given below 
\begin{eqnarray}
\mathrm{pgh}\left( h_{\mu \nu }\right) &=&0,\;\mathrm{pgh}\left( \eta _{\mu
}\right) =1,\;\mathrm{pgh}\left( h^{*\mu \nu }\right) =0=\mathrm{pgh}\left(
\eta ^{*\mu }\right) ,  \label{r112} \\
\mathrm{agh}\left( h_{\mu \nu }\right) &=&0=\mathrm{agh}\left( \eta _{\mu
}\right) ,\;\mathrm{agh}\left( h^{*\mu \nu }\right) =1,\;\mathrm{agh}\left(
\eta ^{*\mu }\right) =2.  \label{r113}
\end{eqnarray}
In agreement with the general line of the antifield-BRST method, the free
BRST differential $s$ for the theory (\ref{r101}) is canonically generated
in the antibracket ($s\cdot =\left( \cdot ,S\right) $) by the solution to
the master equation $\left( S,S\right) =0$, which in our case has the form 
\begin{equation}
S=S^{\mathrm{t}}+S^{\mathrm{h}},  \label{r114}
\end{equation}
where $S^{\mathrm{t}}$ is given by the right-hand side of (\ref{r55}) and 
\begin{equation}
S^{\mathrm{h}}=S_{0}^{\mathrm{PF}}\left[ h_{\mu \nu }\right] +\int
d^{D}x\,h^{*\mu \nu }\partial _{\left( \mu \right. }\eta _{\left. \nu
\right) }.  \label{r115}
\end{equation}

\subsection{First-order deformations: $H\left( \gamma \right) $ and $H\left(
\delta |d\right) $\label{6.2}}

In order to determine the solution to the local first-order deformation
equation (\ref{r62}), we proceed like in Section \ref{5}, namely, we expand
the non-integrated density according to the antighost number as in (\ref{r63}%
) and solve the equivalent tower of equations, given by (\ref{r67}) and (\ref
{r65}--\ref{r66}). It is convenient to split the first-order deformation
into 
\begin{equation}
a=a^{\mathrm{h-h}}+a^{\mathrm{t-t}}+a^{\mathrm{h-t}},  \label{r116}
\end{equation}
where $a^{\mathrm{h-h}}$ denotes the part responsible for the
self-interactions of the Pauli-Fierz field, $a^{\mathrm{t-t}}$ is related to
the deformations of the tensor field $t_{\mu \nu |\alpha \beta }$, and $a^{%
\mathrm{\mathrm{h-t}}}$ signifies the component that describes only the
cross-interactions between $h_{\mu \nu }$ and $t_{\mu \nu |\alpha \beta }$.
Then, $a^{\mathrm{h-h}}$ is completely known (for a detailed analysis, see
for instance~\cite{multi}) 
\begin{equation}
a^{\mathrm{h-h}}=a_{0}^{\mathrm{h-h}}+a_{1}^{\mathrm{h-h}}+a_{2}^{\mathrm{h-h%
}},  \label{r116a}
\end{equation}
where 
\begin{equation}
a_{2}^{\mathrm{h-h}}=\eta ^{*\mu }\eta ^{\alpha }\partial _{\mu }\eta
_{\alpha },  \label{r116b}
\end{equation}
\begin{equation}
a_{1}^{\mathrm{h-h}}=-h^{*\mu \nu }\eta ^{\alpha }\left( \partial _{\mu
}h_{\nu \alpha }+\partial _{\nu }h_{\mu \alpha }-\partial _{\alpha }h_{\mu
\nu }\right) ,  \label{r116c}
\end{equation}
and $a_{0}^{\mathrm{h-h}}$ is the cubic vertex of the Einstein-Hilbert
Lagrangian plus a cosmological term. The piece $a^{\mathrm{t-t}}$ has been
computed in the previous section and is given by the right-hand side of the
formula (\ref{r98}). Inserting (\ref{r116}) in (\ref{r62}) and using the
fact that the first two components already obey the equations 
\begin{equation}
sa^{\mathrm{h-h}}=\partial _{\mu }u^{\mu },\;sa^{\mathrm{t-t}}=\partial
_{\mu }v^{\mu },  \label{r117}
\end{equation}
it follows that only $a^{\mathrm{\mathrm{h-t}}}$ is unknown, being subject
to the equation 
\begin{equation}
sa^{\mathrm{\mathrm{h-t}}}=\partial _{\mu }w^{\mu }.  \label{r118}
\end{equation}
If we develop $a^{\mathrm{\mathrm{h-t}}}$ according to the antighost number 
\begin{equation}
a^{\mathrm{h-t}}=\sum\limits_{k=0}^{I}a_{k}^{\mathrm{h-t}},\;\mathrm{agh}%
\left( a_{k}^{\mathrm{h-t}}\right) =k,\;\mathrm{gh}\left( a_{k}^{\mathrm{h-t}%
}\right) =0,\;\varepsilon \left( a_{k}^{\mathrm{h-t}}\right) =0,
\label{r119}
\end{equation}
(the expansion (\ref{r119}) can be assumed, like in the previous section, to
end at a finite value of the antighost number, once we require that $a_{0}^{%
\mathrm{h-t}}$ is local), then (\ref{r118}) is equivalent to the tower of
equations\footnote{%
The fact that it is possible to replace the equation $\gamma a_{I}^{\mathrm{%
h-t}}=\partial _{\mu }\stackrel{(I)}{w}^{\mu }$ with (\ref{r120}) can be
done like in the proof of Corollary 3.1 from \cite{0402099}.} 
\begin{eqnarray}
\gamma a_{I}^{\mathrm{h-t}} &=&0,  \label{r120} \\
\delta a_{I}^{\mathrm{h-t}}+\gamma a_{I-1}^{\mathrm{h-t}} &=&\partial _{\mu }%
\stackrel{(I-1)}{w}^{\mu },  \label{r121} \\
\delta a_{k}^{\mathrm{h-t}}+\gamma a_{k-1}^{\mathrm{h-t}} &=&\partial _{\mu }%
\stackrel{(k-1)}{w}^{\mu },\;I-1\geq k\geq 1,  \label{r122}
\end{eqnarray}
where $\left( \stackrel{(k)}{w}^{\mu }\right) _{k=\overline{0,I}}$ are some
local currents, with $\mathrm{agh}\left( \stackrel{(k)}{w}^{\mu }\right) =k$.

The equation (\ref{r120}) shows that $a_{I}^{\mathrm{h-t}}\in H\left( \gamma
\right) $, such that on the one hand its solution is unique up to trivial ($%
\gamma $-exact) contributions, $a_{I}^{\mathrm{h-t}}\rightarrow a_{I}^{%
\mathrm{h-t}}+\gamma b_{I}^{\mathrm{h-t}}$, and on the other hand every
purely $\gamma $-exact solution $a_{I}^{\mathrm{h-t}}=\gamma b_{I}^{\mathrm{%
h-t}}$ can be taken to vanish, $a_{I}^{\mathrm{h-t}}=0$. In order to infer
the general solution to this equation, we initially examine the structure of 
$H\left( \gamma \right) $. To this end, from (\ref{r54}) and (\ref{r110a})
we observe that all the antifields 
\begin{equation}
\omega ^{*\Theta }=\left( t^{*\mu \nu |\alpha \beta },h^{*\mu \nu },\eta
^{*\mu \nu |\alpha },\eta ^{*\mu },C^{*\mu \nu }\right)  \label{r123}
\end{equation}
and their spacetime derivatives belong to $H^{0}\left( \gamma \right) $.
Meanwhile, the definition (\ref{r50}) and the first relation in the formula (%
\ref{r110}) yield the most general $\gamma $ -closed (and obviously
non-trivial) objects constructed from the original tensor fields as the
curvature tensor (\ref{curv}), the linearized Riemann tensor (\ref{r107}),
and their derivatives. Consequently, $H^{0}\left( \gamma \right) $ is
spanned by arbitrary polynomials in $\omega ^{*\Theta }$, $F_{\mu \nu
\lambda |\alpha \beta \gamma }$, $K_{\mu \nu |\alpha \beta }$ and their
derivatives. From (\ref{r110}), we observe that the undifferentiated
Pauli-Fierz ghosts $\eta _{\mu }$ and their antisymmetric first-order
derivatives $\partial _{\left[ \mu \right. }\eta _{\left. \nu \right] }$
belong to $H\left( \gamma \right) $, while the symmetric part of their
first-order derivatives is $\gamma $-exact (see the former relation in (\ref
{r110})), and so are all their second- and higher-order derivatives since 
\begin{equation}
\partial _{\alpha }\partial _{\beta }\eta _{\mu }=\frac{1}{2}\gamma \left(
\partial _{\left( \alpha \right. }h_{\left. \beta \right) \mu }-\partial
_{\mu }h_{\alpha \beta }\right) .  \label{r124}
\end{equation}
We have shown in Section \ref{5} that the other set of pure ghost number one
ghosts, related to the tensor field $t_{\mu \nu |\alpha \beta }$, brings no
contribution to $H\left( \gamma \right) $. In conclusion, the presence of
the Pauli-Fierz field enriches the cohomology of $\gamma $, which is no
longer trivial at odd pure ghost numbers, as it happened in the case of the
tensor field $t_{\mu \nu |\alpha \beta }$ alone. Regarding the ghosts of
pure ghost number equal to two, we have seen in the previous section that
the only combinations in $H\left( \gamma \right) $ constructed from them are
polynomials in $C_{\mu \nu }$ and $\partial _{\left[ \mu \right. }C_{\left.
\nu \right] \alpha }$. Thus, the general solution to (\ref{r120}) is
expressed (up to $\gamma $-exact objects) by 
\begin{equation}
a_{I}^{\mathrm{h-t}}=\alpha _{I}^{\mathrm{h-t}}\left( \left[ \omega
^{*\Theta }\right] ,\left[ F_{\mu \nu \lambda |\alpha \beta \gamma }\right]
,\left[ K_{\mu \nu |\alpha \beta }\right] \right) \omega ^{I}\left( \eta
_{\mu },\partial _{\left[ \mu \right. }\eta _{\left. \nu \right] },C_{\mu
\nu },\partial _{\left[ \mu \right. }C_{\left. \nu \right] \alpha }\right) ,
\label{r125}
\end{equation}
for $I>0$, where the $\gamma $-invariant coefficients $\alpha _{I}^{\mathrm{%
h-t}}$ are subject to the conditions $\mathrm{agh}\left( \alpha _{I}^{%
\mathrm{h-t}}\right) =I$ and $\mathrm{pgh}\left( \alpha _{I}^{\mathrm{h-t}%
}\right) =0$, while the symbol $\omega ^{I}$ stands for a generic notation
of the elements with pure ghost number equal to $I$ of a basis of
polynomials in the corresponding ghosts and their antisymmetric first-order
derivatives. In addition, every term in $a_{I}^{\mathrm{h-t}}$ must contain
at least one element from each of the two theories in order to provide
effective cross-interactions. As they have a bounded number of derivatives,
the quantities $\alpha _{I}^{\mathrm{h-t}}$ are in fact polynomials in the
antifields, in the curvature tensor (\ref{curv}), in the linearized Riemann
tensor, and in all their derivatives. They represent the most general
non-trivial elements from $H\left( \gamma \right) $ at pure ghost number
zero and will again be called ``invariant polynomials'' (for the larger free
gauge theory (\ref{r101}), subject to the gauge symmetries (\ref{r8}) and (%
\ref{r103})).

Substituting the solution (\ref{r125}) into the next equation, namely (\ref
{r121}), and taking into account the definitions (\ref{r50}--\ref{r53}) and (%
\ref{r110}--\ref{r111}), we obtain that a necessary condition for the
equation (\ref{r121}) to possess (non-trivial) solutions with respect to $%
a_{I-1}^{\mathrm{h-t}}$ for $I>0$ is that the invariant polynomials $\alpha
_{I}^{\mathrm{h-t}}$ appearing in (\ref{r125}) are non-trivial elements from 
$H_{I}\left( \delta |d\right) $, $\delta \alpha _{I}^{\mathrm{h-t}}=\partial
_{\mu }k^{\mu }$. Taking into account the fact that the maximum Cauchy order
of the free gauge theory (\ref{r101}) is equal to three, we have that~\cite
{gen1},~\cite{gen2} 
\begin{equation}
H_{k}\left( \delta |d\right) =0,\;k>3.  \label{r126}
\end{equation}
Meantime, it remains valid the result that if the invariant polynomial $%
\alpha _{k}^{\mathrm{h-t}}$ is trivial in $H_{k}\left( \delta |d\right) $
for $k\geq 3$, then it can be chosen to be trivial also in $H_{k}^{\mathrm{%
inv}}\left( \delta |d\right) $\footnote{%
The proof can be realized in the same manner like Theorem 4.1 from \cite
{0402099}, with the precaution to include in an appropriate manner the
dependence on the Pauli-Fierz sector.}, which combined with (\ref{r126})
allows us to state that 
\begin{equation}
H_{k}^{\mathrm{inv}}\left( \delta |d\right) =0,\;k>3,  \label{r126a}
\end{equation}
where $H_{k}^{\mathrm{inv}}\left( \delta |d\right) $ denotes, just like
before, the local cohomology group of the Koszul-Tate differential at
antighost number $k$ in the space of invariant polynomials. On account of
the definitions (\ref{r53}) and (\ref{r111}), we are able to identify the
non-trivial representatives of $\left( H_{k}\left( \delta |d\right) \right)
_{k\geq 2}$, as well as of $\left( H_{k}^{\mathrm{inv}}\left( \delta
|d\right) \right) _{k\geq 2}$, under the form 
\begin{equation}
\begin{array}{cc}
\mathrm{agh} & 
\begin{array}{l}
\mathrm{non-trivial\;representatives} \\ 
\mathrm{spanning\;}H_{k}\left( \delta |d\right) \;\mathrm{and}\;H_{k}^{%
\mathrm{inv}}\left( \delta |d\right)
\end{array}
\\ 
k>3 & \mathrm{none} \\ 
k=3 & C^{*\mu \nu } \\ 
k=2 & \eta ^{*\mu \nu |\alpha },\eta ^{*\mu }
\end{array}
.  \label{tabledelta1}
\end{equation}
We will exclude, as we did before, all non-trivial elements from $H\left(
\delta |d\right) $ and $H^{\mathrm{inv}}\left( \delta |d\right) $ at
strictly positive antighost numbers that involve the spacetime co-ordinates,
as they would result in interactions with broken Poincar\'{e} invariance. As
for the cohomological group $H_{1}\left( \delta |d\right) $, its
determination is a difficult task, but we will solve the deformation
equations without explicitly computing it.

Like in the case of the tensor field $t_{\mu \nu |\alpha \beta }$ alone, the
cohomology groups $H_{k}\left( \delta |d\right) $ and $H_{k}^{\mathrm{inv}%
}\left( \delta |d\right) $ at strictly positive antighost numbers give us
information on the obstructions to remove the antifields from the
first-order deformation. As a consequence of the result (\ref{r126a}), we
can eliminate all the terms with $k>3$ from the expansion (\ref{r119}) by
adding to it only trivial pieces, and thus work with $I\leq 3$. This can be
done in principle like in Section 5 from \cite{0402099}, up to the following
observations: (1) the cohomological spaces $\left( H^{2l+1}\left( \gamma
\right) \right) _{l\geq 0}$ are no longer trivial; (2) the operator $\bar{D}$
should be extended to the Pauli-Fierz ghost sector like in the Appendix A.1
from~\cite{multi}. The last representative of (\ref{r119}) is of the type (%
\ref{r125}), with the corresponding invariant polynomials necessarily
non-trivial in $H_{I}^{\mathrm{inv}}\left( \delta |d\right) $ for $I=2,3$,
and respectively in $H_{1}\left( \delta |d\right) $ for $I=1$.

\subsection{The case $I=3$\label{6.3}}

In view of the above considerations we can assume that the expansion (\ref
{r119}) stops at antighost number three ($I=3$) 
\begin{equation}
a^{\mathrm{h-t}}=a_{0}^{\mathrm{h-t}}+a_{1}^{\mathrm{h-t}}+a_{2}^{\mathrm{h-t%
}}+a_{3}^{\mathrm{h-t}},  \label{r127}
\end{equation}
where $a_{3}^{\mathrm{h-t}}$ is of the form (\ref{r125}) for $I=3$. At this
point we enforce the assumption on the maximum derivative order of the
corresponding $a_{0}^{\mathrm{h-t}}$ to be equal to two. Using the result
that the most general representative of $H_{3}^{\mathrm{inv}}\left( \delta
|d\right) $ is the undifferentiated antifield $C^{*\alpha \beta }$ (see (\ref
{tabledelta1}) for $k=3$) and that the elements of pure ghost number three
that fulfill the condition on the maximum derivative order are given by 
\begin{equation}
\left( \eta _{\mu }\eta _{\nu }\eta _{\rho },\eta _{\mu }\eta _{\nu
}\partial _{\left[ \rho \right. }\eta _{\left. \lambda \right] },C_{\mu \nu
}\eta _{\rho },C_{\mu \nu }\partial _{\left[ \rho \right. }\eta _{\left.
\lambda \right] },\partial _{\left[ \mu \right. }C_{\left. \nu \right] \rho
}\eta _{\lambda }\right) ,  \label{r128}
\end{equation}
we can write down that the general solution to the equation (\ref{r120}) for 
$I=3$ like 
\begin{eqnarray}
a_{3}^{\mathrm{h-t}} &=&C^{*\alpha \beta }\left( f_{1\alpha \beta }^{\mu \nu
\rho }\eta _{\mu }\eta _{\nu }\eta _{\rho }+f_{2\alpha \beta }^{\mu \nu \rho
\lambda }\eta _{\mu }\eta _{\nu }\partial _{\left[ \rho \right. }\eta
_{\left. \lambda \right] }+g_{1\alpha \beta }^{\mu \nu \rho }C_{\mu \nu
}\eta _{\rho }\right.  \nonumber \\
&&\left. +g_{2\alpha \beta }^{\mu \nu \rho \lambda }C_{\mu \nu }\partial _{\left[
\rho \right. }\eta _{\left. \lambda \right] }+g_{3\alpha \beta }^{\mu \nu
\rho \lambda }\partial _{\left[ \mu \right. }C_{\left. \nu \right] \rho
}\eta _{\lambda }\right) +\gamma b_{3},  \label{r130}
\end{eqnarray}
where all the coefficients of the type $f$ and $g$ are required to be
non-derivative constants. Combining this result with the symmetries of the
various coefficients due to the corresponding symmetries of the antifield
and of the ghosts, we remain with the following independent possibilities in 
$D\geq 5$ spacetime dimensions: 
\begin{equation}
a_{3}^{\mathrm{h-t}}=a_{3}^{(1)\mathrm{h-t}}+a_{3}^{(2)\mathrm{h-t}%
}+a_{3}^{(3)\mathrm{h-t}},  \label{r131a}
\end{equation}
where

\noindent in $D=5$%
\begin{equation}
a_{3}^{(1)\mathrm{h-t}}=\varepsilon ^{\alpha \beta \mu \nu \rho }C_{\alpha
\beta }^{*}\left( c_{1}\eta _{\mu }\eta _{\nu }\eta _{\rho }+d_{1}C_{\mu \nu
}\eta _{\rho }\right) +\gamma b_{3}^{(1)};  \label{r132}
\end{equation}
in $D=6$%
\begin{eqnarray}
a_{3}^{(2)\mathrm{h-t}} &=&\varepsilon ^{\alpha \beta \mu \nu \rho \lambda
}C_{\alpha \beta }^{*}\left( c_{2}\eta _{\mu }\eta _{\nu }\partial _{\left[
\rho \right. }\eta _{\left. \lambda \right] }\right.  \nonumber \\
&&\left. +d_{2}C_{\mu \nu }\partial _{\left[ \rho \right. }\eta _{\left.
\lambda \right] }+d_{3}\partial _{\left[ \mu \right. }C_{\left. \nu \right]
\rho }\eta _{\lambda }\right) +\gamma b_{3}^{(2)};  \label{r133}
\end{eqnarray}
in all $D\geq 5$\footnote{%
Another possible term in $a_{3}^{(3)\mathrm{h-t}}$ would be $d_{7}C^{*\alpha
\beta }\partial _{\left[ \rho \right. }C_{\left. \alpha \right]
}^{\;\;\;\rho }\eta _{\beta }$, but it is trivial since it can be written
like $\gamma \left( -\frac{d_{7}}{3}C^{*\alpha \beta }\eta _{\alpha \rho
|}^{\;\;\;\;\;\rho }\eta _{\beta }\right) $, and thus we have discarded it
from $a_{3}^{\mathrm{h-t}}$ by putting $d_{7}=0$.} 
\begin{eqnarray}
a_{3}^{(3)\mathrm{h-t}} &=&C^{*\alpha \beta }\left( c_{3}\eta _{\alpha }\eta
^{\rho }\partial _{\left[ \beta \right. }\eta _{\left. \rho \right]
}+d_{4}C_{\alpha }^{\;\;\rho }\partial _{\left[ \rho \right. }\eta _{\left.
\beta \right] }\right.  \nonumber \\
&&\left. +d_{5}\partial _{\left[ \alpha \right. }C_{\left. \beta \right]
\rho }\eta ^{\rho }+d_{6}\partial _{\left[ \rho \right. }C_{\left. \alpha
\right] \beta }\eta ^{\rho }\right) +\gamma b_{3}^{(3)}.  \label{r134}
\end{eqnarray}
In the above all $c_{m}$ and $d_{n}$ are real constants. Obviously, since $%
a_{3}^{\mathrm{h-t}}$ is subject to the equation (\ref{r121}) for $I=3$ and
the components (\ref{r132}--\ref{r134}) are mutually independent, it follows
that each of them must separately fulfill such an equation, i.e., 
\begin{equation}
\delta a_{3}^{(i)\mathrm{h-t}}=-\gamma a_{2}^{(i)\mathrm{h-t}}+\partial
_{\mu }w^{(i)\mu },\;i=1,2,3.  \label{r135}
\end{equation}
By computing the action of $\delta $ on $\left( a_{3}^{(i)\mathrm{h-t}%
}\right) _{i=1,2,3}$ and using the definitions (\ref{r51}) and (\ref{r110}),
we infer that none of them can be written like in the right-hand side of (%
\ref{r135}), no matter what $\left( b_{3}^{(i)}\right) _{i=1,2,3}$ we take
in the right-hand side of (\ref{r132}--\ref{r134}), such that we must set
all the nine constants equal to zero 
\begin{equation}
c_{m}=0,\;m=1,2,3,\;d_{n}=0,\;n=1,2,3,4,5,6,  \label{r136}
\end{equation}
and so $a_{3}^{\mathrm{h-t}}=0$.

\subsection{The case $I=2$\label{6.4}}

We pass to the next eligible value ($I=2$) and write that 
\begin{equation}
a^{\mathrm{h-t}}=a_{0}^{\mathrm{h-t}}+a_{1}^{\mathrm{h-t}}+a_{2}^{\mathrm{h-t%
}}.  \label{r137}
\end{equation}
Repeating the reasoning developed in the above, we obtain that $a_{2}^{%
\mathrm{h-t}}$ is, up to trivial, $\gamma $-exact contributions, of the form
(\ref{r125}) for $I=2$, with the elements of pure ghost number two obeying
the assumption on the maximum number of derivatives from the corresponding $%
a_{0}^{\mathrm{h-t}}$ being equal to two expressed by 
\begin{equation}
\left( \eta _{\mu }\eta _{\nu },\eta _{\mu }\partial _{\left[ \nu \right.
}\eta _{\left. \rho \right] },C_{\mu \nu },\partial _{\left[ \mu \right.
}C_{\left. \nu \right] \rho }\right) .  \label{r137a}
\end{equation}
Using the fact that the general representative of $H_{2}^{\mathrm{inv}%
}\left( \delta |d\right) $ is spanned in this situation by the
undifferentiated antifields $\eta ^{*\alpha \beta |\gamma }$ and $\eta
^{*\alpha }$ (see (\ref{tabledelta1}) for $k=2$), to which we add the
requirement that $a_{2}^{\mathrm{h-t}}$ comprises only terms that
effectively mix the ghost/antifield sectors of the starting free theories,
and combining these with , we obtain that 
\begin{eqnarray}
a_{2}^{\mathrm{h-t}} &=&\eta ^{*\alpha \beta |\gamma }\left( g_{1\alpha
\beta \gamma }^{\mu \nu }\eta _{\mu }\eta _{\nu }+g_{2\alpha \beta \gamma
}^{\mu \nu \rho }\eta _{\mu }\partial _{\left[ \nu \right. }\eta _{\left.
\rho \right] }\right)  \label{r138} \\
&&+\eta ^{*\alpha }\left( g_{3\alpha }^{\mu \nu }C_{\mu \nu }+g_{4\alpha
}^{\mu \nu \rho }\partial _{\left[ \mu \right. }C_{\left. \nu \right] \rho
}\right) +\gamma b_{2},  \nonumber
\end{eqnarray}
where the coefficients denoted by $g$ are imposed to be non-derivative
constants. Taking into account the identity $\eta ^{*\left[ \alpha \beta
|\gamma \right] }\equiv 0$ and the hypothesis that we work only in $D\geq 5$
spacetime dimensions, we arrive at\footnote{%
The possibility $c^{\prime \prime \prime }\eta ^{*\alpha }\partial _{\left[
\alpha \right. }C_{\left. \nu \right] }^{\;\;\;\nu }$ was excluded from $%
a_{2}^{\mathrm{h-t}}$ as it is trivial, being equal to $\gamma \left( -\frac{%
c^{\prime \prime \prime }}{3}\eta ^{*\alpha }\eta _{\alpha \nu
|}^{\;\;\;\;\;\nu }\right) $, such that it can be removed from $a_{2}^{%
\mathrm{h-t}}$ by choosing $c^{\prime \prime \prime }=0$.} 
\begin{equation}
a_{2}^{\mathrm{h-t}}=\frac{c^{\prime }}{2}\eta ^{*\alpha \beta |\mu
}\partial _{\left[ \alpha \right. }\eta _{\left. \beta \right] }\eta _{\mu }+%
\frac{c^{\prime \prime }}{2}\eta _{\;\;\;\;\;\;\;\;\beta }^{*\alpha \beta
|}\partial _{\left[ \alpha \right. }\eta _{\left. \mu \right] }\eta ^{\mu
}+\gamma b_{2}.  \label{r139}
\end{equation}
We will analyse these terms separately. The first one leads to non-vanishing
components of antighost number one and respectively zero as solutions to the
equations 
\begin{equation}
\delta a_{2}^{\prime \mathrm{h-t}}+\gamma a_{1}^{\prime \mathrm{h-t}%
}=\partial _{\mu }\stackrel{(1)}{w^{\prime }}^{\mu },\;\delta a_{1}^{\prime 
\mathrm{h-t}}+\gamma a_{0}^{\prime \mathrm{h-t}}=\partial _{\mu }\stackrel{%
(0)}{w^{\prime }}^{\mu },  \label{r140}
\end{equation}
where we made the notation 
\begin{equation}
a_{2}^{\prime \mathrm{h-t}}=\frac{c^{\prime }}{2}\eta ^{*\alpha \beta |\mu
}\partial _{\left[ \alpha \right. }\eta _{\left. \beta \right] }\eta _{\mu }.
\label{r140a}
\end{equation}
Indeed, straightforward calculations output 
\begin{eqnarray}
a_{1}^{\prime \mathrm{h-t}} &=&\frac{c^{\prime }}{2}t^{*\mu \nu |\alpha
\beta }\left( \left( \partial _{\mu }h_{\nu \alpha }-\partial _{\nu }h_{\mu
\alpha }\right) \eta _{\beta }+\left( \partial _{\alpha }h_{\beta \mu
}-\partial _{\beta }h_{\alpha \mu }\right) \eta _{\nu }\right.  \nonumber \\
&&\left. -\left( \partial _{\mu }h_{\nu \beta }-\partial _{\nu }h_{\mu \beta
}\right) \eta _{\alpha }-\left( \partial _{\alpha }h_{\beta \nu }-\partial
_{\beta }h_{\alpha \nu }\right) \eta _{\mu }\right) ,  \label{r141}
\end{eqnarray}
\begin{equation}
a_{0}^{\prime \mathrm{h-t}}=\frac{c^{\prime }}{8}T^{\mu \nu |\alpha \beta
}\left( h_{\mu \alpha }h_{\nu \beta }-h_{\mu \beta }h_{\nu \alpha }\right) ,
\label{r142}
\end{equation}
where the tensor $T^{\mu \nu |\alpha \beta }$ is given in (\ref{r16}). In
consequence, we obtained a possible form of the first-order deformation for
the cross-interactions between the Pauli-Fierz theory and the tensor field $%
t_{\mu \nu |\alpha \beta }$ like 
\begin{equation}
a^{\prime \mathrm{h-t}}=a_{0}^{\prime \mathrm{h-t}}+a_{1}^{\prime \mathrm{h-t%
}}+a_{2}^{\prime \mathrm{h-t}},  \label{r143}
\end{equation}
where the quantities in the right-hand side of (\ref{r143}) are expressed by
(\ref{r140a}--\ref{r142}). However, $a^{\prime \mathrm{h-t}}$ is trivial in
the context of the overall non-integrated density $a^{\mathrm{h-t}}$ of the
first-order deformation in the sense that it is in a trivial class of the
local cohomology of the free BRST differential $H^{0}\left( s|d\right) $.
Indeed, one can check that it can be put in a $s$-exact modulo $d$ form 
\begin{eqnarray}
a^{\prime \mathrm{h-t}} &=&c^{\prime }s\left( \frac{1}{3}C^{*\mu \nu }\eta
_{\mu }\eta _{\nu }-\frac{1}{2}\eta ^{*\alpha \beta |\mu }\left( h_{\alpha
\mu }\eta _{\beta }-h_{\beta \mu }\eta _{\alpha }\right) \right.  \nonumber
\\
&&\left. +\frac{1}{2}t^{*\mu \nu |\alpha \beta }\left( h_{\mu \alpha }h_{\nu
\beta }-h_{\mu \beta }h_{\nu \alpha }\right) \right) +\partial _{\mu }l^{\mu
},  \label{r144}
\end{eqnarray}
and so it can be eliminated from $a^{\mathrm{h-t}}$ by setting 
\begin{equation}
c^{\prime }=0.  \label{r145}
\end{equation}

The second piece in (\ref{r139}), which is clearly non-trivial, appears to
be more interesting. Indeed, let us fix the trivial ($\gamma $-exact)
contribution from the right-hand side of (\ref{r139}) to 
\begin{equation}
b_{2}=\frac{c^{\prime \prime }}{2}\eta _{\;\;\;\;\;\;\;\;\beta }^{*\alpha
\beta |}h_{\alpha \gamma }\eta ^{\gamma },  \label{r145a}
\end{equation}
which is equivalent to starting from 
\begin{equation}
a_{2}^{\prime \prime \mathrm{h-t}}=c^{\prime \prime }\eta
_{\;\;\;\;\;\;\;\;\beta }^{*\alpha \beta |}\left( \partial _{\alpha }\eta
_{\mu }\right) \eta ^{\mu }.  \label{r146}
\end{equation}
Then, it yields the component of antighost one as solution to the equation $%
\delta a_{2}^{\prime \prime \mathrm{h-t}}+\gamma a_{1}^{\prime \prime 
\mathrm{h-t}}=\partial _{\mu }\stackrel{(1)}{w^{\prime \prime }}^{\mu }$ in
the form 
\begin{equation}
a_{1}^{\prime \prime \mathrm{h-t}}=2c^{\prime \prime }t^{*\mu \alpha }\left(
\partial _{\mu }h_{\alpha \lambda }+\partial _{\alpha }h_{\mu \lambda
}-\partial _{\lambda }h_{\mu \alpha }\right) \eta ^{\lambda },  \label{r147}
\end{equation}
where the notation $t^{*\mu \alpha }$ is explained in (\ref{r44a}). Next, we
pass to the equation 
\begin{equation}
\delta a_{1}^{\prime \prime \mathrm{h-t}}+\gamma a_{0}^{\prime \prime 
\mathrm{h-t}}=\partial _{\mu }\stackrel{(0)}{w^{\prime \prime }}^{\mu },
\label{r148a}
\end{equation}
where 
\begin{equation}
\delta a_{1}^{\prime \prime \mathrm{h-t}}=-\frac{c^{\prime \prime }}{2}%
T^{\mu \alpha }\left( \partial _{\mu }h_{\alpha \lambda }+\partial _{\alpha
}h_{\mu \lambda }-\partial _{\lambda }h_{\mu \alpha }\right) \eta ^{\lambda
},  \label{r149}
\end{equation}
with $T^{\mu \alpha }$ given in (\ref{r16a}). In the sequel we will show
that there are no solutions to (\ref{r148a}). Our procedure goes as follows.
Suppose that there exist solutions $a_{0}^{\prime \prime \mathrm{h-t}}$ to
the equation (\ref{r148a}). Using the formula (\ref{r149}), it follows that
such an $a_{0}^{\prime \prime \mathrm{h-t}}$ must be linear in the tensor
field $t_{\mu \nu |\alpha \beta }$, quadratic in the Pauli-Fierz field, and
second-order in the derivatives. Integrating by parts in the corresponding
functional constructed from $a_{0}^{\prime \prime \mathrm{h-t}}$ allows us
to move the derivatives such as to act only on the Pauli-Fierz fields, and
therefore to work with 
\begin{equation}
a_{0}^{\prime \prime \mathrm{h-t}}=c^{\prime \prime }t^{\mu \nu |\alpha
\beta }a_{\mu \nu |\alpha \beta }^{\mathrm{lin}}\left( h\partial \partial
h,\partial h\partial h\right) ,  \label{r163}
\end{equation}
where the above notation signifies that $a_{\mu \nu |\alpha \beta }^{\mathrm{%
lin}}$ is a linear combination of the generic polynomials between
parentheses (with the mixed symmetry of the tensor field $t_{\mu \nu |\alpha
\beta }$). By direct computation we get that 
\begin{equation}
\gamma a_{0}^{\prime \prime \mathrm{h-t}}=\partial ^{\mu }\left( 4c^{\prime
\prime }\eta ^{\alpha \beta |\nu }a_{\mu \nu |\alpha \beta }^{\mathrm{lin}%
}\right) -4c^{\prime \prime }\eta ^{\alpha \beta |\nu }\partial ^{\mu
}a_{\mu \nu |\alpha \beta }^{\mathrm{lin}}+c^{\prime \prime }t^{\mu \nu
|\alpha \beta }\gamma a_{\mu \nu |\alpha \beta }^{\mathrm{lin}},
\label{r165z}
\end{equation}
where 
\begin{equation}
\gamma a_{\mu \nu |\alpha \beta }^{\mathrm{lin}}=\bar{a}_{\mu \nu |\alpha
\beta }^{\mathrm{lin}}\left( h\partial \partial \partial \eta ,\partial
h\partial \partial \eta ,\partial \partial h\partial \eta \right) ,
\label{r166}
\end{equation}
with $\eta $ a generic notation for the Pauli-Fierz ghost $\eta _{\mu }$. As 
$\delta a_{1}^{\prime \prime \mathrm{h-t}}$ contains no ghosts from the $%
t_{\mu \nu |\alpha \beta }$-sector, we require that $\gamma a_{0}^{\prime
\prime \mathrm{h-t}}$ obeys the property 
\begin{equation}
\partial ^{\mu }a_{\mu \nu |\alpha \beta }^{\mathrm{lin}}\left( h\partial
\partial h,\partial h\partial h\right) =0,  \label{r164}
\end{equation}
such that 
\begin{equation}
\gamma a_{0}^{\prime \prime \mathrm{h-t}}=\partial ^{\mu }\left( 4c^{\prime
\prime }\eta ^{\alpha \beta |\nu }a_{\mu \nu |\alpha \beta }^{\mathrm{lin}%
}\right) +c^{\prime \prime }t^{\mu \nu |\alpha \beta }\gamma a_{\mu \nu
|\alpha \beta }^{\mathrm{lin}}.  \label{r165}
\end{equation}
Simple calculations in (\ref{r149}) give 
\begin{equation}
\delta a_{1}^{\prime \prime \mathrm{h-t}}=\partial _{\mu }p^{\mu }+c^{\prime
\prime }t^{\mu \nu |\alpha \beta }b_{\mu \nu |\alpha \beta }^{\mathrm{lin}%
}\left( \partial h\partial \partial \eta ,\partial \partial h\partial \eta
,\eta \partial \partial \partial h\right) .  \label{r167}
\end{equation}
Inserting (\ref{r165}--\ref{r167}) in (\ref{r148a}) and observing that only $%
b_{\mu \nu |\alpha \beta }^{\mathrm{lin}}$ contains terms that are
third-order in the derivatives of the Pauli-Fierz fields, we conclude that
the existence of $a_{0}^{\prime \prime \mathrm{h-t}}$ is completely dictated
by the behaviour of $b_{\mu \nu |\alpha \beta }^{\mathrm{lin}}$. More
precisely, $a_{0}^{\prime \prime \mathrm{h-t}}$ exists if and only if the
part of the type $\eta \partial \partial \partial h$ from $b_{\mu \nu
|\alpha \beta }^{\mathrm{lin}}$ vanishes identically and/or can be written
like the $\delta $-variation of something like $\partial h^{*}t\eta $.
Direct computation produces the part from $b_{\mu \nu |\alpha \beta }^{%
\mathrm{lin}}$ of order three in the derivatives of the Pauli-Fierz fields
in the form 
\begin{eqnarray}
b_{\mu \nu |\alpha \beta }^{\mathrm{lin}}\left( \eta \partial \partial
\partial h\right) &\sim &c^{\prime \prime }\eta ^{\lambda }\partial
_{\lambda }\left( \sigma _{\beta \nu }\left( \partial _{\mu }\partial ^{\rho
}h_{\rho \alpha }+\partial _{\alpha }\partial ^{\rho }h_{\rho \mu }-\Box
h_{\alpha \mu }-\partial _{\alpha }\partial _{\mu }h\right) \right. 
\nonumber \\
&&-\frac{1}{2}\sigma _{\beta \nu }\sigma _{\alpha \mu }\left( \partial
^{\rho }\partial ^{\gamma }h_{\rho \gamma }-\Box h\right) +\partial _{\beta
}\partial _{\nu }h_{\alpha \mu }  \nonumber \\
&&+\left( \alpha \longleftrightarrow \beta ,\mu \longleftrightarrow \nu
\right)  \nonumber \\
&&-\left( \beta \longleftrightarrow \alpha ,\mu \rightarrow \mu ,\nu
\rightarrow \nu \right)  \nonumber \\
&&\left. -\left( \mu \longleftrightarrow \nu ,\alpha \rightarrow \alpha
,\beta \rightarrow \beta \right) \right) ,  \label{r168}
\end{eqnarray}
and it neither vanishes identically nor is proportional with $\delta \left(
\partial _{\lambda }h_{\alpha \mu }^{*}\right) $, as it can be observed from
the expression (\ref{r106}) of the functions that define the field equations
for the Pauli-Fierz field. The rest of the terms from (\ref{r168}) are
obtained from the first ones by making the indicated index-changes. In
conclusion, we must also take 
\begin{equation}
c^{\prime \prime }=0  \label{r168a}
\end{equation}
in (\ref{r146}), so $a_{2}^{\mathrm{h-t}}=0$.

\subsection{The case $I=1$\label{6.5}}

Now, we analyse the next possibility, namely $I=1$ in (\ref{r119}) 
\begin{equation}
a^{\mathrm{h-t}}=a_{0}^{\mathrm{h-t}}+a_{1}^{\mathrm{h-t}},  \label{r152}
\end{equation}
where $a_{1}^{\mathrm{h-t}}$ must be searched among the non-trivial
solutions to the equation $\gamma a_{1}^{\mathrm{h-t}}=0$, which are offered
by 
\begin{equation}
a_{1}^{\mathrm{h-t}}=\alpha _{1}^{\mathrm{h-t}}\left( \left[ t^{*\mu \nu
|\alpha \beta }\right] ,\left[ h^{*\mu \nu }\right] ,\left[ F_{\mu \nu
\lambda |\alpha \beta \gamma }\right] ,\left[ K_{\mu \nu |\alpha \beta
}\right] \right) \omega ^{1}\left( \eta _{\mu },\partial _{\left[ \mu
\right. }\eta _{\left. \nu \right] }\right) ,  \label{r152a}
\end{equation}
where the elements of pure ghost number one are 
\begin{equation}
\left( \eta _{\mu },\partial _{\left[ \mu \right. }\eta _{\left. \nu \right]
}\right) .  \label{r153}
\end{equation}
On the one hand, the assumption on the maximum derivative order of the
interacting Lagrangian being equal to two prevents the coefficients $\alpha
_{1}^{\mathrm{h-t}}$ to depend on either the curvature tensors or their
spacetime derivatives. On the other hand, $a_{1}^{\mathrm{h-t}}$ can involve
only the antifields $t^{*\mu \nu |\alpha \beta }$ and their spacetime
derivatives, because otherwise, as $\omega ^{1}$ includes only the
Pauli-Fierz ghosts, it would not lead to cross-interactions between the
fields $t_{\mu \nu |\alpha \beta }$ and $h_{\mu \nu }$. Moving in addition
the derivatives from these antifields such as to act only on the elements (%
\ref{r153}) from $a_{1}^{\mathrm{h-t}}$ and relying again on the assumption
on the maximum derivative order, we eventually remain with one possibility%
\footnote{%
The identity $t^{*\left[ \mu \nu |\alpha \right] \beta }=0$ forbids the
appearance of solutions proportional with Levi-Civita symbols in any $D\geq
5 $ dimension.} (up to $\gamma $-exact quantities) 
\begin{eqnarray}
a_{1}^{\mathrm{h-t}} &\sim &t^{*\mu \nu |\alpha \beta }\left( \sigma _{\mu
\alpha }\partial _{\left[ \nu \right. }\eta _{\left. \beta \right] }-\sigma
_{\mu \beta }\partial _{\left[ \nu \right. }\eta _{\left. \alpha \right]
}+\sigma _{\nu \beta }\partial _{\left[ \mu \right. }\eta _{\left. \alpha
\right] }-\sigma _{\nu \alpha }\partial _{\left[ \mu \right. }\eta _{\left.
\beta \right] }\right)  \nonumber \\
&=&4t^{*\nu \beta }\partial _{\left[ \nu \right. }\eta _{\left. \beta
\right] }\equiv 0,  \label{r154}
\end{eqnarray}
which vanishes identically due to the symmetry property in (\ref{r44a}) of
the simple trace of the antifield $t^{*\mu \nu |\alpha \beta }$.

\subsection{The case $I=0$\label{6.6}}

As $a_{1}^{\mathrm{h-t}}$ in (\ref{r154}) vanishes, we remain with one more
case, namely where $a^{\mathrm{h-t}}$ reduces to its antighost number zero
piece 
\begin{equation}
a^{\mathrm{h-t}}=a_{0}^{\mathrm{h-t}}\left( \left[ t_{\mu \nu |\alpha \beta
}\right] ,\left[ h_{\mu \nu }\right] \right) ,  \label{r155}
\end{equation}
which is subject to the equation 
\begin{equation}
\gamma a_{0}^{\mathrm{h-t}}=\partial _{\mu }w^{\mu }.  \label{r156}
\end{equation}
As we have discussed in Section \ref{5}, there are two types of solutions to
(\ref{r156}). The first one corresponds to $w^{\mu }=0$ and is given by
arbitrary polynomials that mix the curvature tensor (\ref{curv}) and its
spacetime derivatives with the linearized Riemann tensor (\ref{r107}) and
its derivatives, which are however excluded from the condition on the
maximum derivative order of $a_{0}^{\mathrm{h-t}}$ (their derivative order
is at least four). The second one is associated with $w^{\mu }\neq 0$, being
understood that we discard the divergence-like solutions $a_{0}^{\mathrm{h-t}%
}=\partial _{\mu }z^{\mu }$ and preserve the maximum derivative-order
restriction. Denoting the Euler-Lagrange derivatives of $a_{0}^{\mathrm{h-t}%
} $ by $B^{\mu \nu |\alpha \beta }\equiv \delta a_{0}^{\mathrm{h-t}}/\delta
t_{\mu \nu |\alpha \beta }$ and respectively by $D^{\mu \nu }=\delta a_{0}^{%
\mathrm{h-t}}/\delta h_{\mu \nu }$, and using the formula (\ref{r50})
together with the first definition in (\ref{r110}), the equation (\ref{r156}%
) implies that 
\begin{equation}
\partial _{\mu }B^{\mu \nu |\alpha \beta }=0,\;\partial _{\mu }D^{\mu \nu
}=0.  \label{r157}
\end{equation}
The tensors $B^{\mu \nu |\alpha \beta }$ and $D^{\mu \nu }$ are imposed to
contain at most two derivatives and to have the mixed symmetry of $t_{\mu
\nu |\alpha \beta }$ and respectively of $h_{\mu \nu }$. Meanwhile, they
must yield a Lagrangian $a_{0}^{\mathrm{h-t}}$ that effectively couples the
two sorts of fields, so $B^{\mu \nu |\alpha \beta }$ and $D^{\mu \nu }$
effectively depend on $h_{\mu \nu }$ and respectively on $t_{\mu \nu |\alpha
\beta }$. According to the considerations from Section \ref{2} and
Subsection \ref{6.1} (see the formulas (\ref{r40a}--\ref{r40b}) and (\ref
{r109c}--\ref{r109d})), the solutions to the equations (\ref{r157}) are of
the type\footnote{%
The solutions involving the constant tensors $B^{\mu \nu |\alpha \beta }\sim
\left( \sigma ^{\mu \alpha }\sigma ^{\nu \beta }-\sigma ^{\mu \beta }\sigma
^{\nu \alpha }\right) $ and $D^{\mu \nu }\sim \sigma ^{\mu \nu }$ give
cosmological terms and have already been considered in the above. They are
not eligible anyway in the present context, which exclusively focuses on the
cross-interactions between the two sorts of fields.} 
\begin{equation}
\frac{\delta a_{0}^{\mathrm{h-t}}}{\delta t_{\mu \nu |\alpha \beta }}\equiv
B^{\mu \nu |\alpha \beta }=\partial _{\rho }\partial _{\gamma }\tilde{\Phi}%
^{\mu \nu \rho |\alpha \beta \gamma },\;\frac{\delta a_{0}^{\mathrm{h-t}}}{%
\delta h_{\mu \nu }}\equiv D^{\mu \nu }=\partial _{\alpha }\partial _{\beta }%
\tilde{\Phi}^{\mu \alpha |\nu \beta },  \label{r157a}
\end{equation}
where $\tilde{\Phi}^{\mu \nu \rho |\alpha \beta \gamma }$ and $\tilde{\Phi}%
^{\mu \alpha |\nu \beta }$ depend only on the undifferentiated fields $%
h_{\mu \nu }$ and $t_{\mu \nu |\alpha \beta }$ (otherwise, the
corresponding $a_{0}^{\mathrm{h-t}}$ would be more than second-order in the
derivatives), with $\tilde{\Phi}^{\mu \nu \rho |\alpha \beta \gamma }$
having the mixed symmetry of the curvature tensor $F^{\mu \nu \rho |\alpha
\beta \gamma }$ and $\tilde{\Phi}^{\mu \alpha |\nu \beta }$ that of the
linearized Riemann tensor. From now on we proceed along the line employed in
the Subsection \ref{5.3}. In view of this, we introduce a derivation on the
algebra of non-integrated densities depending on $t_{\mu \nu |\alpha \beta }$%
, $h_{\mu \nu }$ and on their derivatives, that counts the powers of the
fields and their derivatives 
\begin{eqnarray}
\bar{N} &=&\sum\limits_{n\geq 0}\left( \left( \partial _{\mu _{1}}\cdots
\partial _{\mu _{n}}t_{\mu \nu |\alpha \beta }\right) \frac{\partial }{%
\partial \left( \partial _{\mu _{1}}\cdots \partial _{\mu _{n}}t_{\mu \nu
|\alpha \beta }\right) }\right.  \nonumber \\
&&\left. +\left( \partial _{\mu _{1}}\cdots \partial _{\mu _{n}}h_{\mu \nu
}\right) \frac{\partial }{\partial \left( \partial _{\mu _{1}}\cdots
\partial _{\mu _{n}}h_{\mu \nu }\right) }\right) ,  \label{r157b}
\end{eqnarray}
and observe that the action of $\bar{N}$ on an arbitrary non-integrated
density $\bar{u}\left( \left[ t_{\mu \nu |\alpha \beta }\right] ,\left[
h_{\mu \nu }\right] \right) $ is 
\begin{equation}
\bar{N}\bar{u}=t_{\mu \nu |\alpha \beta }\frac{\delta \bar{u}}{\delta t_{\mu
\nu |\alpha \beta }}+h_{\mu \nu }\frac{\delta \bar{u}}{\delta h_{\mu \nu }}%
+\partial _{\mu }r^{\mu },  \label{r157c}
\end{equation}
where $\delta \bar{u}/\delta t_{\mu \nu |\alpha \beta }$ and $\delta \bar{u}%
/\delta h_{\mu \nu }$ denote the variational derivatives of $\bar{u}$. In
the case where $\bar{u}$ is a homogeneous polynomial of order $p>0$ in the
fields and their derivatives, we have that $\bar{N}\bar{u}=p\bar{u}$, and so 
\begin{equation}
\bar{u}=\frac{1}{p}\left( t_{\mu \nu |\alpha \beta }\frac{\delta \bar{u}}{%
\delta t_{\mu \nu |\alpha \beta }}+h_{\mu \nu }\frac{\delta \bar{u}}{\delta
h_{\mu \nu }}\right) +\partial _{\mu }\left( \frac{1}{p}r^{\mu }\right) .
\label{r157d}
\end{equation}
As $a_{0}^{\mathrm{h-t}}$ can always be decomposed as a sum of homogeneous
polynomials of various orders, it is enough to analyze the equation (\ref
{r156}) for a fixed value of $p$. Putting $\bar{u}=a_{0}^{\mathrm{h-t}}$ in (%
\ref{r157d}) and inserting (\ref{r157a}) in the associated relation, we can
write 
\begin{equation}
a_{0}^{\mathrm{h-t}}=\frac{1}{p}\left( t_{\mu \nu |\alpha \beta }\partial
_{\rho }\partial _{\gamma }\tilde{\Phi}^{\mu \nu \rho |\alpha \beta \gamma
}+h_{\mu \nu }\partial _{\alpha }\partial _{\beta }\tilde{\Phi}^{\mu \alpha
|\nu \beta }\right) +\partial _{\mu }\bar{r}^{\mu }.  \label{r157e}
\end{equation}
Integrating twice by parts in (\ref{r157e}) and recalling the mixed
symmetries of $\tilde{\Phi}^{\mu \nu \rho |\alpha \beta \gamma }$ and $%
\tilde{\Phi}^{\mu \alpha |\nu \beta }$, we infer that 
\begin{equation}
a_{0}^{\mathrm{h-t}}=k_{1}F_{\mu \nu \rho |\alpha \beta \gamma }\tilde{\Phi}%
^{\mu \nu \rho |\alpha \beta \gamma }+k_{2}K_{\mu \alpha |\nu \beta }\tilde{%
\Phi}^{\mu \alpha |\nu \beta }+\partial _{\mu }\bar{l}^{\mu },  \label{r157f}
\end{equation}
with $k_{1}=1/9p$ and $k_{2}=-1/2p$. By computing the action of $\gamma $ on
(\ref{r157f}) and following a reasoning similar to that applied between the
formulas (\ref{r97h}) and (\ref{r97p}), we obtain that $p=2$ and 
\begin{equation}
a_{0}^{\mathrm{h-t}}=k^{\prime }T^{\mu \alpha }h_{\mu \alpha }.  \label{r158}
\end{equation}
As the above $a_{0}^{\mathrm{h-t}}$ vanishes on the stationary surface (\ref
{r15}) of the field equations for the tensor $t_{\mu \nu |\alpha \beta }$,
it is trivial in $H^{0}\left( s|d\right) $. Indeed, by direct computation we
have that 
\begin{equation}
a_{0}^{\mathrm{h-t}}=s\left( 2k^{\prime }\left( 2t^{*\mu \alpha }h_{\mu
\alpha }+\eta _{\;\;\;\;\;\;\beta }^{*\alpha \beta |}\eta _{\alpha }\right)
\right) +\partial _{\mu }\left( -8k^{\prime }t^{*\mu \alpha }\eta _{\alpha
}\right) ,  \label{r159}
\end{equation}
so it can be removed from the first-order deformation by choosing 
\begin{equation}
k^{\prime }=0.  \label{r160}
\end{equation}

Putting together the results contained in this section, we can state that $%
S_{1}^{\mathrm{h-t}}=0$ and so 
\begin{equation}
S_{1}=S_{1}^{\mathrm{h-h}}+S_{1}^{\mathrm{t-t}},  \label{r160a}
\end{equation}
where $S_{1}^{\mathrm{h-h}}$ is the first-order deformation of the solution
to the master equation for the Pauli-Fierz theory and $S_{1}^{\mathrm{t-t}}$
is given in the right-hand side of (\ref{r98}). The consistency of $S_{1}$
at the second order in the coupling constant is governed by the equation (%
\ref{r61}), where $\left( S_{1}^{\mathrm{h-h}},S_{1}^{\mathrm{t-t}}\right)
=0=\left( S_{1}^{\mathrm{t-t}},S_{1}^{\mathrm{t-t}}\right) $, and thus we
have that $S_{2}^{\mathrm{t-t}}=0=S_{2}^{\mathrm{h-t}}$, while $S_{2}^{%
\mathrm{h-h}}$ is highly non-trivial and is known to describe the quartic
vertex of the Einstein-Hilbert action, as well as the second-order
contributions to the gauge transformations and to the associated non-abelian
gauge algebra. The vanishing of $S_{1}^{\mathrm{h-t}}$ and $S_{2}^{\mathrm{%
h-t}}$ further leads, via the equations that stipulate the higher-order
deformation equations, to the result that actually 
\begin{equation}
S_{k}^{\mathrm{h-t}}=0,\;k\geq 1.  \label{r160c}
\end{equation}

The main conclusion of this section is that, under the general conditions of
smoothness, locality, Lorentz covariance and Poincar\'{e} invariance of the
deformations, combined with the requirement that the interacting Lagrangian
is at most second-order derivative, there are no consistent, non-trivial
cross-couplings between the Pauli-Fierz field and the massless tensor field
with the mixed symmetry of the Riemann tensor. The only pieces that can be
added to the action (\ref{r101}) are given by the cosmological term for the
tensor $t_{\mu \nu |\alpha \beta }$ and, naturally, by the self-interactions
of the Pauli-Fierz field, which produce the Einstein-Hilbert action,
invariant under the diffeomorphisms.

\section{Interactions with matter fields\label{7}}

In the final part of this paper we show that the massless tensor field with
the mixed symmetry of the Riemann tensor cannot be coupled in a consistent,
non-trivial manner to any matter theory such that the matter fields gain
gauge transformations. Indeed, let us consider a generic matter theory 
\begin{equation}
S^{\mathrm{matt}}\left[ y^{i}\right] =\int d^{D}x\mathcal{L}\left( \left[
y^{i}\right] \right) ,  \label{r169}
\end{equation}
where the fields $y^{i}$ are assumed to have no non-trivial gauge
symmetries. In this situation, the BRST differential for the action written
as the sum between (\ref{r1}) and (\ref{r169}) acts on the BRST generators
according to (\ref{r50}--\ref{r53}) and respectively to 
\begin{equation}
\gamma y^{i}=0,\;\gamma y_{i}^{*}=0,\;\delta y^{i}=0,\;\delta y_{i}^{*}=-%
\frac{\delta ^{L}\mathcal{L}}{\delta y^{i}},  \label{r170}
\end{equation}
where 
\begin{equation}
\mathrm{pgh}\left( y^{i}\right) =0=\mathrm{pgh}\left( y_{i}^{*}\right) ,\;%
\mathrm{agh}\left( y^{i}\right) =0,\;\mathrm{agh}\left( y_{i}^{*}\right) =1,
\label{r171}
\end{equation}
and $y_{i}^{*}$ denote the antifields of the matter fields. The presence of
the matter theory simply adds to $H\left( \gamma \right) $ discussed in
Subsection \ref{hgama} the dependence on $y^{i}$, $y_{i}^{*}$ and their
spacetime derivatives, which lie at pure ghost number zero, $\left[
y^{i}\right] ,\left[ y_{i}^{*}\right] \in H^{0}\left( \gamma \right) $, and
therefore we still have that $H^{2l+1}\left( \gamma \right) =0$. From (\ref
{r170}) it is clear that the cross-interactions between the tensor field $%
t_{\mu \nu |\alpha \beta }$ and the matter fields $y^{i}$ at the first order
in the coupling constant can be produced just by a first-order deformation
of the master equation that stops at antighost number one, $a^{\mathrm{t-matt%
}}=a_{1}^{\mathrm{t-matt}}+a_{0}^{\mathrm{t-matt}}$, where $\gamma a_{1}^{%
\mathrm{t-matt}}=0$. However, as $H^{1}\left( \gamma \right) $ is trivial,
this fact implies that $a_{1}^{\mathrm{t-matt}}$ is trivial and consequently
the matter fields cannot gain gauge invariance. We remain with the sole
possibility that $a^{\mathrm{t-matt}}=a_{0}^{\mathrm{t-matt}}$, with $\gamma
a_{0}^{\mathrm{t-matt}}=\partial _{\mu }q^{\mu }$, whose solutions, once we
add the restriction on the maximum derivative order of the cross-couplings
being equal to two, are spanned by polynomials that are simultaneously
linear in the curvature tensor (\ref{curv}) and of any order in the
undifferentiated matter fields.

\section{Conclusion}

The general conclusion of this paper is that the powerful reformulation of
interactions in gauge theories in terms of the local BRST cohomology reveal
that the massless tensor field with the mixed symmetry of the Riemann tensor
admits no consistent self-interactions and, in the meantime, cannot be
coupled in a consistent, non-trivial manner to the massless spin-two field,
described in the free limit by the Pauli-Fierz theory. We also argued that
the attempt to couple such a mixed symmetry-type tensor to purely matter
theories produces no gauge transformations with respect to the matter field
sector. Our analysis was constantly based on the assumptions that the
resulting deformations are smooth, local, Lorentz-covariant and
Poincar\'{e}-invariant and on the natural requirement that the maximum
derivative order of the interacting Lagrangian is equal to two. It is
possible that the relaxation of the last condition yields non-trivial,
consistent interactions, at least with the massless spin-two fields, in
which case the first-order formulation~\cite{zinov2, zinov3} of such an
tensor field would probably be a happier starting point.

\section*{Acknowledgment}

This work has been supported from a type A grant with the Romanian National
Council for Academic Scientific Research and the Romanian Ministry of
Education, Research and Youth. The authors thank Nicolas Boulanger for
valuable suggestions and comments.

\end{document}